\documentclass{aa}
\usepackage{epsfig}
\usepackage{natbib}
\usepackage{lscape}
\usepackage{longtable}
\newcommand\nodata{ ~$\cdots$~ }
\begin{document}

\title{WIYN Open Cluster Study. XXVI. Improved kinematic membership
and spectroscopy of IC~2391
\thanks{Based on observations collected
at the European Southern Observatory, Chile (Program IDs: 072.D-0107 and 
074.D-0096)}
\thanks{Tables 1, 3, 5, and 6 are only available in electronic form at the
CDS via anonymous ftp to cdsarc.u-strasbg.fr (130.79.128.5)
or via http://cdsweb.u-strasbg.fr/cgi-bin/qcat?J/A+A/ }
}
\titlerunning{IC~2391}

\author{I. Platais\inst{1}
\and
C. Melo\inst{2,3}
\and
J.-C. Mermilliod\inst{4}
\and
V. Kozhurina-Platais\inst{5}
\and
J. P. Fulbright\inst{1}
\and
R. A. M\'{e}ndez\inst{2}
\and
M. Altmann\inst{2}
\and	    
J. Sperauskas\inst{6}
   }
\institute{
Department of Physics and Astronomy, Johns Hopkins University, 
3400 North Charles Street, Baltimore, MD 21218, USA
\and
Departamento de Astronom\'{i}a, Universidad de Chile, Casilla 36-D, Santiago, Chile
\and
European Southern Observatory, Casilla 19001, Santiago 19, Chile
\and
Laboratoire d'Astrophysique de l'Ecole polytechnique f\'{e}d\'{e}rale de Lausanne, 1290 Chavannes-des-Bois, Switzerland
\and
Space Telescope Science Institute, 3700 San Martin Drive, Baltimore,
MD 21218, USA
\and
Vilnius University Observatory, Ciurlionio 29, Vilnius 2009, Lithuania
}

\offprints{I. Platais\\email:~imants@pha.jhu.edu}
\date{Received 2 June 2006/ Accepted 21 August 2006}

\abstract
{Young open clusters provide important
clues to the interface between the main sequence and pre-main-sequence
phases of stellar evolution. The young and nearby open cluster IC~2391 
is well-suited to studies of these two evolutionary phases.}
{We establish a bona fide set of cluster members and then 
analyze this set in terms of binary frequency, projected rotational
velocities, [Fe/H], and lithium abundance. In the
wake of the Hipparcos distance controversy for the Pleiades, we compare 
the main-sequence fitting distance modulus to the Hipparcos mean
parallax for IC~2391.}
{We have obtained new proper motions for 6991 stars 
down to $V$$\sim$13-16 over a $\sim$9-deg$^2$ area of the sky comprising
IC~2391. The precision of proper motions, $\sigma_\mu=1.7$ mas~yr$^{-1}$, 
allowed us to calculate reliable membership probabilities.  
We also obtained precise radial velocity and $v \sin i$ measurements with 
\textsc{Coravel} and FEROS for 76 probable cluster members. The cluster's
mean radial velocity is $+14.8\pm$0.7 km~s$^{-1}$.
The FEROS high-resolution spectra were used to determine both
the [Fe/H] abundance in the four main sequence dwarfs of IC~2391 and the Li
abundance in 47 stars.  In addition, new $BV$ CCD photometry was 
obtained for the majority of probable cluster members.}
{The proper-motion survey covers a 6 times larger sky
area than the prior targeted searches for cluster members in IC~2391.
A total of 66 stars are considered bona fide cluster members
down to a mass equivalent to 0.5M$_{\sun}$.
A quarter of them have been newly identified with many in the
F2-K5 spectral range, which is crucial for a main-sequence fit.
We find a mean [Fe/H] value of $+0.06\pm0.06$, when a solar abundance
of $\log \epsilon$(Fe)=7.45 is adopted.
The main sequence fitting yields a distance modulus that is 0.19 mag larger 
than that derived from Hipparcos parallaxes; thus this offset 
nearly has the size of a similar offset found for the Pleiades.
The Li abundance pattern is similar to the earlier findings 
and is typical for a 40 Myr old open cluster.}
{A variety of new data on the probable members of IC~2391 improve essentially
all observational parameters of this young open cluster.}

\maketitle

\section{Introduction}

IC~2391 is a young ($\sim$35~Myr) and nearby ($d\sim$150~pc) open cluster
located in Vela ($\ell=270\degr$, $b=-7\degr$). Its proximity 
is very appealing for any detailed studies of intrinsically faint
low-mass stars and brown dwarfs \citep{bar04}.
The significance of IC~2391 is clearly demonstrated by a large number
of literature references over the last decade in the SIMBAD Astronomical 
Database: $\sim$200 publications where the cluster has been mentioned
or studied.  One problem facing the researchers of IC~2391
is the scarcity of confirmed
cluster members. For a long time, the known cluster membership was confined to 
merely $\sim$20 stars, all brighter than $V$$\sim$11 \citep{hog60}.
Then, \citet{sta89} reported on the proper-motion study of
883 stars over a $48\arcmin\times41\arcmin$ area.  Proper motions, 
$BV\!RI$ photometry, and high-resolution spectroscopy together
yielded a list of ten additional probable cluster members down 
to $V=14$. This list was substantially extended by using the ROSAT
imaging data \citep{pat93,pat96,sim98} to take
advantage of the known strong X-ray activity among the young 
G-K-M spectral type stars. In the follow-up spectroscopic study,
\citet{sta97} confirmed the cluster membership of 23
X-ray selected stars down to $V$$\sim$15, using the radial velocity,
Li line, and H$_\alpha$ appearance as membership criteria.
\citet{dod04} attempted to identify more cluster members by
mining USNO-B and 2MASS catalogs. From 185 astrometrically selected
possible cluster members, a total of 35 stars are brighter than
$R=15$. However, a disturbingly small fraction of these stars
($\sim$20\%) are common with the Patten \& Simon (1993) list in
the same magnitude range and spatial coverage. The latest search
for cluster members in the central $30\arcmin\times30\arcmin$ region
of IC~2391 by the XMM-Newton X-ray observatory resulted in nine 
relatively faint possible new members \citep{mar05}.

As indicated above, proper motions have been used as a kinematic
membership discriminator for IC~2391. However,  
only the study by \citet{kin79} provides precise relative proper
motions ($\sigma=0.9$ mas yr$^{-1}$) down to $V$$\sim$12 
over a $1\fdg7\times0\fdg9$ area.
In this study, from a total of 232 stars about 
40 have proper motions consistent with membership in IC~2391.
No formal membership probabilities are calculated, apparently owing 
to the sparseness of the cluster. 
Another way to ascertain the membership status, independent of
any assumptions on the astrophysical properties of probable
cluster members, is to use radial velocities.
There is a rich literature on this subject for IC~2391, e.g.,
\citet{fei61}, \citet{bus65}, \citet{per69a}, \citet{vhoo72}, 
\citet{lev88}, \citet{sta97}, \citet{bar99}.
Nearly 100 stars have had their
radial velocities measured, many of them several times.
In many cases, however, the precision of the radial velocities was
low, especially for the early and very late type stars, thereby largely
precluding assignment of a reliable membership status to them.

Often IC~2391 is considered along with IC~2602, because both have very
similar properties and are separated spatially only by $\sim$50~pc, thus
suggesting a common origin.  Their absolute proper motions, however, 
differ significantly.  The projected total velocity in the
tangential plane for IC~2391 is 33.8 mas yr$^{-1}$, whereas
for IC~2602 it is only 20.5 mas yr$^{-1}$ \citep{rob99}. 
A much larger tangential velocity of IC~2391 considerably
increases the reliability of membership probabilities drawn from
proper motions, since a smaller fraction of field stars are
expected to share the motion of the cluster. This and the limited 
precision of our proper motions are one of the main reasons
for selecting IC~2391 as the subject of this study. 

\citet{ran01} spectroscopically analyzed $\sim$50 X-ray selected
candidate members in IC~2391 and IC~2602. From the analysis of 8
Fe~I lines in four stars, the mean metallicity of IC~2391 was derived 
to be [Fe/H]=$-0.03\pm0.07$. In this study, Li abundance was
obtained for 32 possible members of IC~2391, covering a wide 
range of $T_{\rm eff}$ -- from 3500 to 6600 K. It was found that stars 
warmer than $\sim$5800 K or more massive than $\sim$1~M$_\odot$ show no 
significant signs of Li depletion. For cooler late-G to
early-K stars, the pattern of Li abundances in IC~2391 and the
Pleiades is similar, although hinting that in this $T_{\rm eff}$ range
Li is less depleted in IC~2391, as one would expect from the age 
difference. A more detailed analysis of Li abundance in IC~2391 is 
hindered by the small number of stars in the \citet{ran01} sample.

Young open clusters appear to have stars with a broad range of rotational
rates \citep{her05}. That is also confirmed by the observed
rotational rates for late-type stars \citep{pat96}
and the $v\sin i$ distribution in IC~2391 \citep{sta97}.
From the standpoint of Li abundances, X-ray luminosities, and
stellar evolution, it is vital to identify fast rotators in
the enlarged sample of cluster stars.

In some aspects, the level of our understanding of the open cluster IC~2391 is
similar to NGC~2451A, which was recently studied by \citet{pla01} as one of
the WIYN Open Cluster Study (WOCS) targets.
The lack of comprehensive astrometric cluster membership prompted
us to include IC~2391 among the WOCS clusters. Following the WOCS
strategy \citep{mat00}, we derived new proper motions and 
calculated the cluster membership probabilities. For many probable
cluster members, high-resolution spectroscopy served to measure the
radial velocities, projected rotational velocities $v\sin i$, 
Li abundance, and equivalent width of H$_\alpha$. A few carefully
selected cluster stars are used to obtain metallicity [Fe/H].
New CCD photometry is used to construct reliable color-magnitude diagram
and perform the isochrone fit.  

\section{Astrometric reductions and cluster membership}

A total of four 8$\times$10 inch photographic
 plates (scale $=55\farcs1$ mm$^{-1}$), taken
with the 51~cm double astrograph of Cesco Observatory in El Leoncito,
Argentina, were used for astrometry.  Two of these
visual-bandpass plates (103a-G emulsion and OG-515 filter) were
obtained in 1967.29, the other two in 1996.14. An
objective wire-grating was used to produce diffraction images for
all stars brighter than $V$$\sim$13. Each first-epoch plate contains
two exposures: a 30~min and an offset 1~min exposure. 

Our target stars were drawn from the COSMOS/UKST Object Catalog
\citep{yen92}. In this catalog the object brightness is
given in $B_{\rm J}$ magnitudes as derived in the natural photographic system 
(IIa-J emulsion and GG-395 filter) of the UK 1.2 m Schmidt Telescope 
at Siding Spring, Australia \citep{bla82}. 
Due to the scan-time limitations set by the
measuring machine, the sample selection required an optimization. All stars
down to $B_{\rm J}=13.0$ were chosen in a $3\fdg5\times2\fdg7$
rectangle centered on $\alpha=8^{\rm h} 40^{\rm m}$ and
$\delta=-52\degr 53\arcmin$ (equinox J2000.0). A sub-sample
of fainter stars at $B_{\rm J}=14.6$ in the same area served as
anonymous astrometric reference stars. Then, within this rectangle
all additional stars down to $B_{\rm J}=16.2$ were selected in a circle 
with the radius of $0\fdg8$ centered on $\alpha=8^{\rm h} 42\fm5$ 
and $\delta=-53\degr$.  

Altogether, our initial sample included
over 7,000 stars. All measurable images of these stars were digitized
with the Yale 2020G PDS microdensitometer in a fine-raster,
object-by-object mode. The image positions were determined using the
Yale Image Centering routine \citep{lee83}, which
includes a two-dimensional Gaussian fit. 

The positions and proper motions were calculated using the standard
SPM (Southern Proper Motion program) astrometric reductions, 
described in detail by \citet{gir98} and \citet{pla98}. 
Owing to the relatively small 9-deg$^2$ field, only linear and
quadratic plate-tilt terms were used in the proper-motion plate model. The
standard error of proper motions was estimated to range from
1.4 to 2.1 mas yr$^{-1}$, depending on the star's magnitude and hence
on the number of available grating images. The calculated relative
proper motions are free of apparent systematic errors. The
distribution of proper motions or a vector-point diagram (VPD)
is shown in Fig.~1. A visible clumping of proper motions at
$\mu_x=-20$, $\mu_y=+18$ mas yr$^{-1}$ indicates the presence of
IC~2391 members.

The local sample method \citep{koz95} was used to
calculate the cluster membership. In this method, for each target
star a representative sub-sample (a bin) of other stars is formed that
shares the properties of the target, such as the brightness.
We used a wide 10-magnitude sliding brightness bin, which for the
brightest and faintest stars narrows down to 5 magnitudes.
No spatial window was used for this sample of proper motions.
Similar to the case of NGC 2451A \citep{pla01}, a flat
distribution of field stars in VPD was adopted in the vicinity 
of the cluster centroid. The resulting membership probability,
 $P_\mu$, is defined as

\begin{equation}
P_\mu=\frac{\Phi_c}{\Phi_c+\Phi_f},
\end{equation}

\noindent where, $\Phi_c$ is a Gaussian representing the cluster star 
distribution in the VPD
and $\Phi_f$ is the distribution of field stars, both defined within
the same magnitude bin. The following parameters were adopted for
the cluster star distribution: the cluster center in the VPD at
$\mu_{x}^{c}=-20.1$ and $\mu_{y}^{c}=+17.6$ mas yr$^{-1}$; the Gaussian
width $\sigma_{c}=1.7$ mas yr$^{-1}$. A total of 115 stars, all having
membership probability $P_{\mu}\geq5\%$, are listed in 
Table~1. A formal sum of
membership probabilities indicates that in our sample of 6991 stars,
only $\sim$65 are members of IC~2391.

\begin{figure}
  \resizebox{\hsize}{!}{\includegraphics[angle=-90]{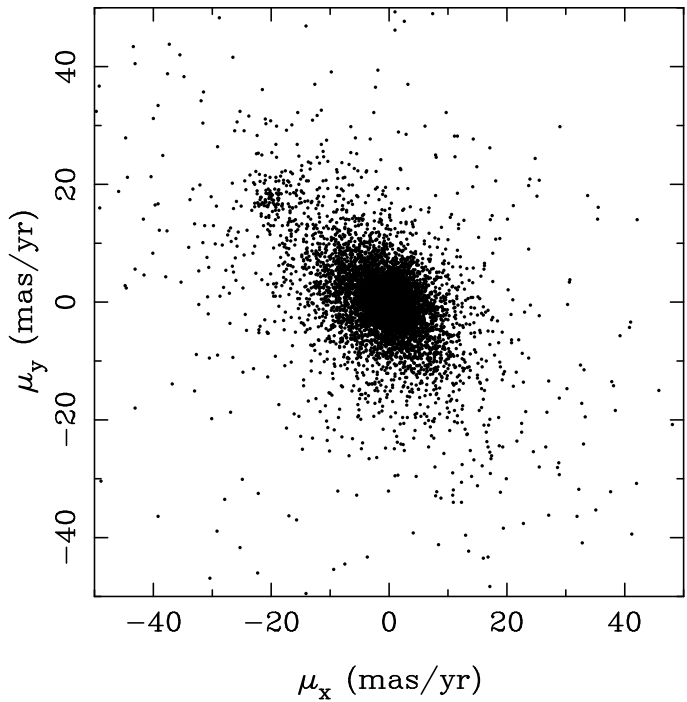}}
\caption{Proper-motion vector point diagram in the area of IC~2391.
A clump of proper motions at $\mu_x=-20.1$ and $\mu_y=+17.6$ mas yr$^{-1}$
indicates the location of the cluster.
}
\label{fig:vpd}
\end{figure}

Finally, precise equatorial coordinates were calculated for all stars,
choosing UCAC2 stars \citep{zac04} as a reference frame.
The coordinate transformation required a quadratic plate model
supplemented with two main cubic distortion terms. The standard
error of that transformation via the least-squares formalism
is $\sim$60~mas. At the epoch and equinox of J2000, the estimated 
average accuracy of the catalog\footnote{The catalog is available at
http://www.astro.yale.edu/astrom/} positions is about 30 mas. We note 
that this study provides precise coordinates for many X-ray selected 
cluster members, which so far have had only approximate coordinates from
\citet{pat96} and \citet{mar05}.

\section{Photometry and CCD reductions}

A few sources providing $U\!BV$ photometry cover mainly the inner area
of IC~2391, e.g., \citet{hog60}, \citet{lyn60}, \citet{per69b}.
In January 1997 we obtained new CCD $BV$ photometry for the
majority of the possible proper-motion members. The observations were 
made at the Cerro Tololo Inter-American Observatory (CTIO) 0.9~m
telescope with the Tektronics $2048\times2048$ CCD chip, which covers
$13\farcm5\times13\farcm5$ on the sky.

As in the case of NGC 2451A \citep{pla01}, each probable cluster
member or else a group of members with mutual separations less than 
$\sim$10$\arcmin$ was observed individually. Only a small fraction
of these stars could be identified on more than one CCD frame in each filter.
In total, 114 CCD frames 
were obtained in $B$ and $V$ filters with exposure time varied from 1 to
150~s, depending on the star's magnitude. A set of \citet{lan92} standards
was taken three times a night over 22-24 January, 1997 -- spanning our
observations of IC~2391.  The twilight sky frames were used to correct 
the pixel-to-pixel sensitivity variations.

All CCD frames were reduced using the IRAF DAOPHOTX photometry package.
The aperture photometry routine PHOT was applied because the target
stars were optimally exposed and well-isolated. The details of
transformation of the instrumental magnitudes into the standard
$BV$ system for this observing run are given in \citet{pla01}.
The final $BV$ magnitudes are believed to be on the standard system
to within 0.02 mag and have a standard error of 0.03 mag in $V$ and
0.02 in $B-V$. It should be noted that the reddest stars may have 
a slightly less accurate photometry since the reddest standard star
has only $B-V=1.18$.

Table~\ref{tab:phot} shows the comparison of our CCD $BV$ photometry with seven 
other sources of $BV$ photometry, mainly photoelectric. This table 
contains the literature reference, the number of common stars $(n)$,
mean $\Delta V$ and $\Delta (B-V)$ in the sense of our CCD 
photometry minus the published one. The errors are standard
deviations from the mean calculated difference in magnitude or color.
All the sources of photometry
are consistent, though on average our CCD $V$-magnitudes 
appear to be fainter by $\sim$0.03 than those from the other sources.

\section{\textsc{Coravel} radial velocities and $v\sin i$}

The radial-velocity observations were made with the photoelectric spectrometer 
\textsc{Coravel} \citep{bar79,may85} on the Danish 1.54~m telescope at 
ESO, La Silla, Chile. They were obtained starting in
March 1984 through April 1996 (when the \textsc{Coravel} was retired)
during the course of regular 
ESO and Danish time runs allocated to the open cluster studies.
In the \textsc{Coravel}-type instruments, the spectrum of a star is 
electro-mechanically correlated, i.e., scanned with an appropriate 
spectral mask in the focal plane.  The output correlation profile 
can be described with a Gaussian (position, depth, width) 
plus a continuum level, readily providing an estimate of radial
velocity and $v\sin i$. The latter is computed following the
techniques described by \citet{ben81,ben84}.
The radial velocities are on the system defined by \citet{udr99}, calibrated 
with high-precision data from the \textsc{ELODIE} spectrograph 
\citep{bar96}.

The initial sample consisted of only 8 stars, all brighter than $B=11.6$
and already thought to be possible members in 1983.  
Three of them (3664=SHJM~7, 3722, 5382) were found to be SB1 
spectroscopic binaries and two were found to be SB2 (389, 4413). 
It should be noted that \textsc{Coravel} can detect a secondary if the
magnitude difference is smaller than $\sim$1.5-1.8 mag.

The list of candidate cluster members was greatly enlarged by the ROSAT
X-ray detections in the area of IC~2391 \citep{pat93,pat96}. A total of
17 such stars ($B<13.8$) were observed once or twice in February 1995 and/or
January-April 1996. Among them, two additional likely 
spectroscopic binaries (5768, 5859) were  detected. 

Two presumably constant stars are common to the \textsc{Coravel} and FEROS 
(see Sect. 5) samples.  The radial velocity difference, in the sense
\textsc{Coravel}$-$FEROS, for star 4362 is +0.09 km~s$^{-1}$ and for
star 4809 is $-0.10$ km~s$^{-1}$. This indicates very good agreement
between the two systems.
There are 13 stars in common with the radial-velocity data obtained in 
1995 at CTIO by \citet{sta97}.  One of them, 5859=VXR~67a, appears to be 
a spectroscopic binary judging from two \textsc{Coravel} observations 
(Table~3).
The spectroscopic binary, suspected by \citet{sta97}, 4549=VXR~30, is
a definite SB1 from the \textsc{Coravel} data.
For the remaining 11 stars, the radial velocity difference 
`\textsc{Coravel}$-$CTIO' is $-0.6\pm0.3$ km~s$^{-1}$, which is in good
agreement with the listed internal and external errors by \citet{sta97}.

Table~3  contains all \textsc{Coravel} heliocentric 
radial velocities and their
estimated standard errors. The last four entries were obtained with
the Lithuanian \textsc{Coravel}-type spectrometer \citep{upg02} at the CASLEO
2.2~m telescope in El Leoncito, Argentina, in February 2002.

Table~\ref{tab:meanrv}  lists the mean radial velocities, 
$v\sin i$, the associated
errors and other parameters. For a star, depending on the structure
of the correlation profile, up to three sets of parameters 
(for components A and B, and a blend) can be given.
The estimated formal standard deviation of the mean radial velocity
is denoted by $\sigma$.  The estimated  mean internal uncertainty
of a radial velocity measurement is denoted by 
$\epsilon=$ max$(\sigma, mean~error)/\sqrt{n}$, where $n$ is
the number of measurements. The internal error of an individual observation
consists of three components as described by \citet{bar79} and \citet{mer89}.
We note that the uncertainty, $\epsilon$, is progressively underestimated at 
increasing $v\sin i$ values \citep{nor96}. 
The ratio, standard deviation $\sigma$ vs. the mean internal uncertainty of
individual measurements, denoted by the E/I, is used to calculate the
probability $P(\chi^2)$ that the scatter is due to random noise.
A star is considered a spectroscopic binary, if $P(\chi^2)\leq0.001$.
In the case of a single measurement, the $P(\chi^2)$ is meaningless and
marked by 9.999.  Additional columns 6, 7, 8, and 9 contain the
number of measurements, $n$, the time span in days covered by the observations,
$\Delta T$, measured $v\sin i$ and its standard error. Binarity status
and the component are indicated in last column.

Finally, we note that the systems of $v\sin i$ from \textsc{Coravel}, FEROS, and
\citet{ran01} are all in excellent agreement for $v\sin i\leq 20$ km~s$^{-1}$
to within $\sim$1 km s$^{-1}$, but can differ by up to $\sim$10 km~s$^{-1}$
for large $v\sin i$ and/or in some cases of spectroscopic binaries.

\section{Spectroscopy with FEROS}

The spectroscopic observations were carried out in two observing runs 
in February 2004 and February 2005 using the high-resolution, two-fiber
($R\sim$50,000)  FEROS echelle spectrograph  \citep{kaufer99} at the ESO
2.2~m telescope in La Silla, Chile. This spectrograph provides
a wavelength range from 360 to 920 nm, covered by 39 echelle orders.
Such a wide wavelength range is essential in the chemical abundance 
studies allowing one to select appropriate unblended metal lines. 
The observations were taken in the Object-Sky mode
in which the target is centered onto the Object fiber, whereas the Sky fiber
collects the light from the sky background.  The reductions of spectra 
were performed using the standard FEROS pipeline, which includes flat-fielding, 
sky-background subtraction, removal of cosmic rays, wavelength calibration, 
and barycentric velocity correction. The pipeline yields a 1-D re-binned
spectrum evenly sampled at 0.03\AA~ steps.

Radial velocities are derived by cross-correlation techniques
using a K0 III spectral type digital binary mask as the template 
\citep{bar79,queloz95}.
The resulting cross-correlation function (CCF) can in most cases be 
approximated by a Gaussian function whose center readily gives the 
radial velocity and the width (Gaussian $\sigma$) related to 
broadening mechanisms such as turbulent motion, gravity pressure, and rotation.

Owing to a relatively high
$S/N$, normally higher than 50 (see Table~5),
the photon noise errors in our radial velocity measurements
typically range between 5-15 m s$^{-1}$.  The final uncertainties in
derived radial velocities are a combination of the photon noise
errors and  the overnight spectrograph
drift due to the changes in the index of air refractivity and
atmospheric pressure. Usually, these shifts are on the order of a few
hundred m~s$^{-1}$ per night.  One way to correct for this
drift is to have the calibration lamp
illuminate a second fiber while the first fiber receives the
stellar light.  Although the FEROS two-fiber configuration allows
for this option \citep{setiawan00}, the subsequent data reduction and
analysis is complicated and not warranted for young stars where
the  radial-velocity jitter exceeds the internal precision
of measurement by a lot. For example, for very active and young T~Tauri
type stars, this jitter is not lower than $\sim$0.6 km~s$^{-1}$ and can be
as high as $\sim$2 km~s$^{-1}$ \citep{melo03}. With age these
effects gradually abate, though the rotational modulation of stellar
active regions can generate a radial velocity scatter up to
50 m~s$^{-1}$ even at the age of Hyades at $\sim$600 Myr \citep{pau04}.
We therefore opted for a simpler approach, as described below.

\begin{figure}
  \resizebox{\hsize}{!}{\includegraphics[angle=-90]{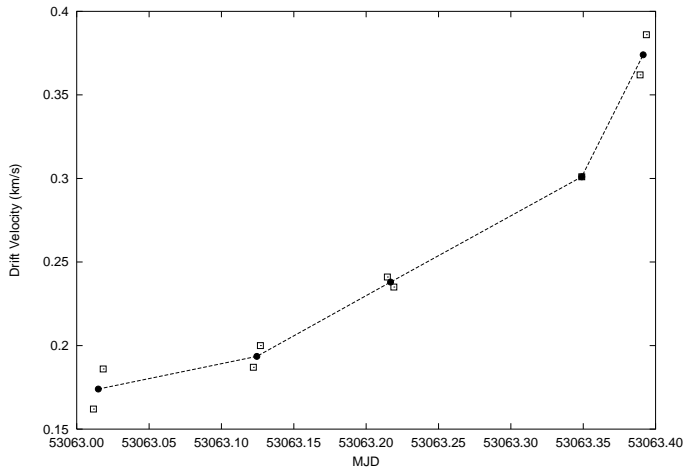}}
\caption{Drift correction for the night of February 27, 2004. Standard stars
are shown as open squares and the mean drift as filled circles.
The dashed line is the adopted drift as a function of time.}
\label{fig:drift}
\end{figure}

In order to correct for this drift, 1-3 radial velocity standards from 
the CORALIE extra-solar planet survey were observed a few times 
during the night. For each standard star, the velocity drift 
was computed as the difference between the CORALIE radial velocity and 
the observed radial velocity. In the case of more than one standard star,
the mean drift was calculated. The radial-velocity corrections for 
the program stars were computed using a linear interpolation between 
each two drift points. A typical drift correction is shown in
Fig.~\ref{fig:drift}.  Over two observing runs, the r.m.s.
of the differences between the corrected radial velocities of standard stars 
and the CORALIE radial velocities is below 20 m s$^{-1}$, indicating
that the procedure works well. One should keep in mind, however, that 
the final uncertainty of radial velocities could be higher, since 
at some level a linear interpolation might not approximate the actual
pattern of a drift accurately. We note that the r.m.s. scatter of the 
drift correction between the FEROS and CORALIE radial velocities is 
$\sim$150 m~s${^-1}$, similar to the value found by \cite{melo01} using 
$\tau$ Cet as a standard.

A few double-lined spectroscopic binaries were found among our target stars
(Table~5, Fig.~\ref{fig:sbs}). Their cluster membership
status is briefly discussed in Sect.~6.4.

\subsection{FEROS $v\sin i$}

The width, $\sigma $, of the cross-correlation function (CCF) is a product
of several broadening mechanisms related to gravity, turbulence,
magnetic fields, effective temperature, metallicity, and rotation. In addition,
the instrumental profile also contributes to the broadening of spectral
lines and, therefore, to the CCF. Thus, in order to correctly measure the
contribution of rotation to the width of the CCF, we should model all
sources of broadening, except what is due to rotation.
The width of FEROS CCF was calibrated as described in \cite{melo01}.
For fast rotators ($v\sin i \ga 30$ km~s$^{-1}$), the final $v\sin i$
was derived as follows. The CCF is fitted by a family of functions
$CCF_{v\sin i}=C-D[g_0\otimes G(v\sin i)]$ which is a convolution
of the CCF of a non-rotating star, $g_0$, in turn approximated by a Gaussian
and the \cite{Gray92} rotational profile
computed for a set of discrete rotational velocities, $G(v\sin i)$. For each
function $CCF_{v\sin i}$, we can find the radial velocity $V_{\rm r}$,
the depth D, and the continuum C by minimizing the quantity
$\chi^2_{v\sin i}$.  Figure~1 in \citet{melo03} illustrates this method.
The uncertainty of the $v\sin i$ measurement is $\sim$1.5 km~s$^{-1}$ for
$v\sin \la 30$ km~s$^{-1}$ and $\sim$10\% of the  $v\sin i$ value
for $v\sin \ga 30$ km~s$^{-1}$. The FEROS measurements of $v\sin i$ 
are listed in Table~5. The averaged $v\sin i$ values
are given in Table~\ref{tab:meanli}. In the case of SB2 spectroscopic
binaries, an average $v\sin i$ is given only for the A-component.

\subsection{[Fe/H] determination}

The FEROS high-resolution spectra are used to determine the [Fe/H] abundance
of IC~2391.  However, other abundance studies of relatively young open 
clusters, such as the Hyades, Pleiades, and the Ursa Major moving group 
\citep{sch06,yon04,pau03}, 
have found that the atmospheres of the G- and K-dwarfs show deviations
from the simple, plane-parallel atmosphere models used in most abundance
studies.  High excitation lines and lines of ionized species are stronger
than predicted by simple model atmospheres, even though the same
models perform adequately for older G- and K- dwarfs like the Sun.
\citet{sch06} suggest that photospheric spots and faculae in young
dwarfs can possibly produce such deviations.

Until the solution to this problem is found, we only perform a
simple abundance analysis on a few stars in IC~2391.  The atmosphere's
problem is minimized in warmer G-dwarfs, hence we chose 
four slowly-rotating G-dwarfs with $T_{\rm eff}$ values ranging 
between 5200~K and 5900~K in the sample.

We adopted the list of iron lines from \citet{ful06}.  The lines in this
list were selected to be those least affected by blending for use in the 
study of metal-rich bulge giants, where the differential analysis was
done relative to Arcturus.
For this study, we used the Sun as the differential standard
and adopt a solar iron abundance of $\log \epsilon$(Fe)$=$7.45.  The lines were measured manually
using the IRAF {\it splot} package, and the measured equivalent widths are given
in Table~6. We measured the solar line equivalent widths from
the Solar Atlas
by \citet{kur84} and a high $S/N$ ($>$200) FEROS sky spectrum. 
We did not use lines stronger than $\sim$120 m\AA.

There is good agreement between the results from the Solar Atlas and the 
FEROS solar spectrum.  If the $gf$-values of the Fe lines are adjusted to
yield a solar Fe abundance of $\log \epsilon$(Fe)$=$7.45 for the equivalent
widths measured in the Solar Atlas, the FEROS sky spectrum yields
an Fe~I abundance of $7.44 \pm 0.06$ (97 lines) and an Fe~II abundance
of $7.42 \pm 0.07$ (5 lines).  Therefore, we believe that the instrumental 
effects in our differential analysis method are negligible.

Due to the aforementioned potential problems with the atmospheres,
we did not use 
spectroscopic indicators to set the stellar parameters, with the exception
of setting the model atmosphere [m/H] value to match the derived [Fe/H]
value.  We used a grid of solar-abundance ratio atmospheres by Fiorella
Castelli\footnote{http://wwwuser.oat.ts.astro.it/castelli/} that include
updated opacity distribution functions and the 2002 version
of the MOOG spectrum synthesis program \citep{moog}.

We set the $T_{\rm eff}$ values as an average from three 
color-to-$T_{\rm eff}$ calibrations for dwarf stars \citep{ram05}, 
adopting the initial [Fe/H]=0, very close to the estimate 
by \citet{ran01}. We used $B-V$,
$V-J$, and $V-K_s$ color indices to obtain $T_{\rm eff}$, where
$J$ and $K_s$ are the 2MASS magnitudes \citep{skr06}. In the temperature
range 4400-5700 K, the common stars between our study and 
\citet{ran01} indicate a nearly perfect match of temperature
scales. If all nine common stars are considered, the r.m.s. scatter of
differences approaches 140 K. It should not be overlooked, however,
that such photometrically-calibrated effective temperatures may be
biased (e.g., cooler), if a star is a binary and/or still on the
pre-main-sequence evolutionary tracks. The \citet{ram05} calibrations do not
consider such effects though their presence in IC~2391 is
undeniable (see Sect. 6.5).
We derived the surface gravities $\log g$ by 
interpolation of the 35-Myr, solar-metallicity isochrone by
\citet{gir02}.  Finally, we used the relationship provided by 
\citet{all04} to set the microturbulence parameter $v_{\rm t}$.

The derived atmosphere parameters and Fe abundances are given in 
Table~\ref{tab:abund}.  The line-by-line abundances show
slight trends with respect to excitation potential and line strength,
but we did not adjust the stellar parameters or remove any high-excitation 
lines from the analysis.  We point out an especially large difference 
(mean of $0.15 \pm 0.06$ dex)
between the abundances derived from Fe~I and Fe~II lines, noting that
by definition these same lines give the same abundance for both species
in the Sun.  This is similar to what was seen in studies of other 
young open clusters, so we only used the results for the
Fe~I lines to derive the mean cluster abundance.

The weighted mean [Fe~I/H] 
value for the four IC~2391 stars is $+0.06 \pm 0.06$ (s.d.). When adopting the
solar abundance of $\log \epsilon$(Fe)$=$7.45, the mean [Fe/H] reported
by \citet{ran01} translates into 
[Fe/H]=$+0.04\pm0.07$, a metallicity estimate nearly identical to
our value. Recently, in the framework of UVES Paranal Observatory Project,
\citet{stu06} obtained elemental abundances for five early-type stars 
in IC~2391.  The weighted mean [Fe/H] value for the two bona fide cluster
members (HD~74275=4522 and SHJM~2=3722) from this study is $+0.10\pm0.07$, 
again in good agreement with our value of [Fe/H]. A third bona fide
member HD~74044=7027 was also included in the \citet{stu06} analysis;
however, it shows signs of being chemically peculiar and, hence, is not
suitable for deriving average elemental abundances in cluster stars.

Abundances derived from Fe~I lines are sensitive to the adopted
$T_{\rm eff}$ values:  a $+100$~K increase in $T_{\rm eff}$ will raise
the [Fe/H] value by about $+0.10$~dex.  Our temperature scale comes only
from photometric colors, so it is sensitive to errors in the photometry,
to color shifts due to unresolved binaries, and to the intrinsic
uncertainties of the fitting functions used to derive the $T_{\rm eff}$
scales (the $\sigma T_{\rm eff}$ values the T(B-V), T(V-J), and T(V-K)
calibrations are 88~K, 62~K, and 50~K, respectively).  For example,
we note that star 3359 is slightly above the isochrone fitting the main
sequence in two out of the three color-magnitude diagrams (see Sect. 6.5).
If this is due to the problem in photometry,
and we have adopted a $T_{\rm eff}$ value that could be lower than
true $T_{\rm eff}$ by $\sim$50~K, then our final [Fe/H] value for
star 3359 is too low by $\sim$0.05~dex.

\subsection{Li abundance}
 
The equivalent width of Li 6708\AA\ is known to be a good youth 
indicator for cool stars (late G to mid-M type stars)
whose interiors are fully convective during the pre-main sequence 
phase \cite[e.g.,][]{martin97}.  For hotter stars (F to mid-G 
corresponding to $B-V=0.4-0.7$), the development of a radiative core 
in the pre-main-sequence phase prevents a rapid Li depletion
in these stars. Thus, for hotter stars, the changes in Li abundance
are insignificant between a few Myr and the Pleiades age \cite[e.g.,][]{sod93}. 

\begin{figure}
  \resizebox{\hsize}{!}{\includegraphics[angle=-90]{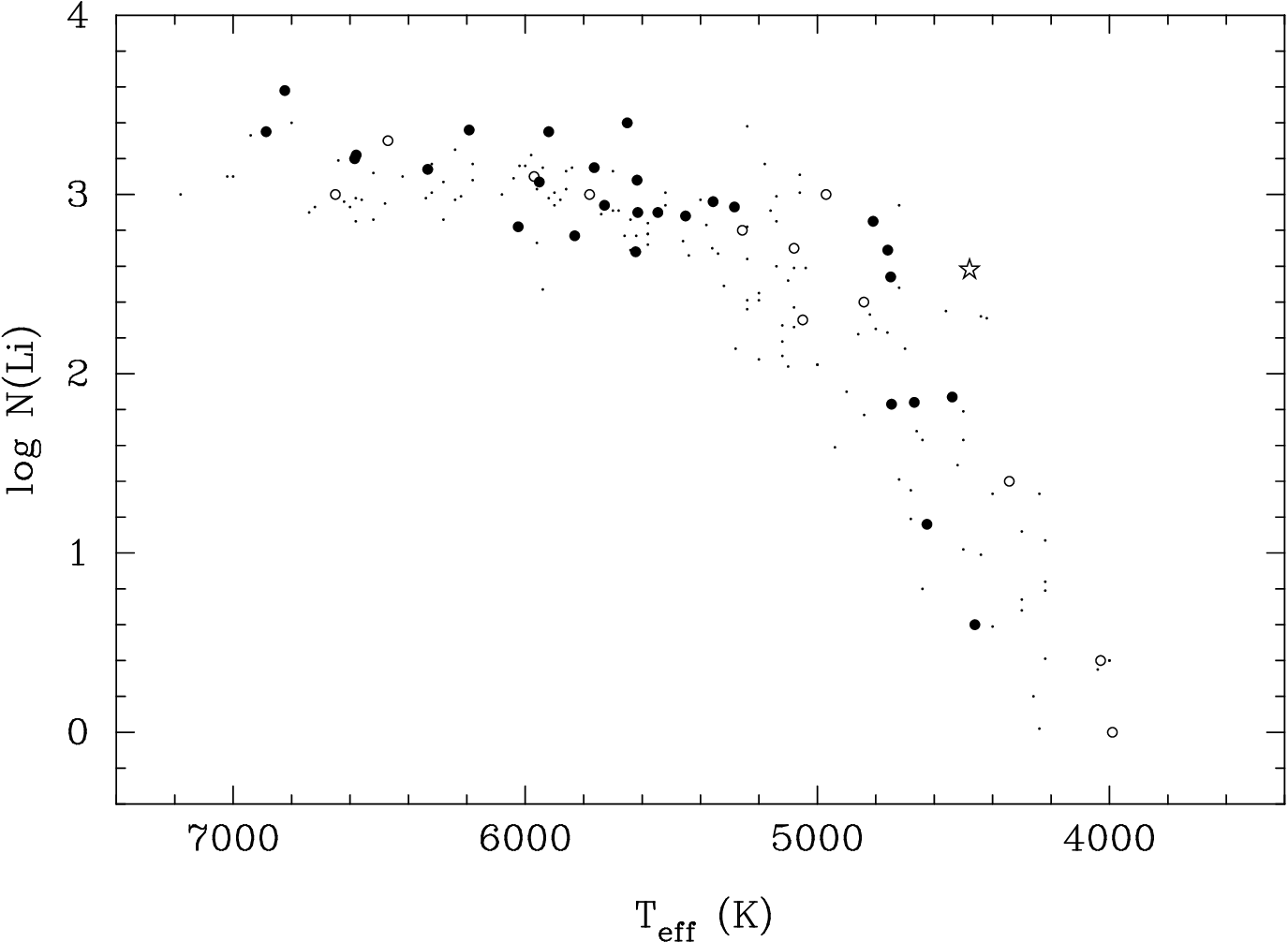}}
\caption{Lithium abundance as a function of $T_{\rm eff}$. The bold
dots denote bona fide members of IC~2391 from Table~\ref{tab:meanli}.
The circles indicate bona fide members having their Li abundance
measured only by \citet{ran01}. The location of uncertain member
1820=SHJM~3 (see Sect. 6.2) is marked by the open star.  
Superimposed dots represent the Pleiades
members from \citet{sod93}. It appears that in the range 
$4600<T_{\rm eff}<4800$ K the cluster stars undergo a rapid phase of  
Li depletion.
}
\label{fig:li}
\end{figure}

The equivalent width of Li 6708\AA\ and H$_\alpha$ was measured
manually using 
the IRAF {\it splot} task. Because in fast rotating stars this Li feature is 
blended with an Fe I line at 6707.441~\AA, we applied an empirical
correction \citep[Sect.~2.2][]{sod93} to the equivalent width of the Li 
feature for all stars with $v\sin i>10$ km~s$^{-1}$. Then, the equivalent
width was converted into Li abundance, $\log$ N(Li), via the grid of
curves of growth tabulated by \citet{sod93}. Effective temperatures
were derived as described in Sect.~5.2. The initial analysis of the
$\log$ N(Li) distribution as a function of $T_{\rm eff}$ showed
a number of stars with abnormally high Li abundance, 
exceeding $\log$ N(Li)$=3.5$ dex. In our FEROS spectra these stars 
have $v\sin i \geq40$ km~s$^{-1}$ and a poorly-defined continuum around
the Li feature.  For such stars, it was decided to adopt the smoothed
adjacent spectrum just outside the expected width of a Li feature
for a continuum.  
The measured equivalent widths, EW(Li), are listed in Table~\ref{tab:meanli}. 
Ambiguous lithium non-detections due to the high rotational 
velocity or binarity are marked by $-99$. 
For stars with $v\sin i<15$ km~s$^{-1}$, the r.m.s. scatter of measured EWs
from two independent measurements is estimated to be $\sim$6 m\AA.
In this range of $v\sin i$, 
there are only three stars, 4362, 4413, and 5859, in common with \citet{ran01}.
The measurements of EW in both studies agree to within $2\sigma$ of the
errors quoted by \citet{ran01}. For bona fide cluster members of IC~2391,
the distribution of lithium abundance as a function of $T_{\rm eff}$ is
shown in Fig.~\ref{fig:li}.

From the FEROS spectra we also estimated an equivalent width of H$_{\alpha}$.
Eight relatively cool stars in Table~\ref{tab:meanli} show H$_{\alpha}$ 
in emission and therefore have negative EW values.

\section{Results}

One of the goals of this paper is to assemble a list of bona fide members
of IC~2391 and their basic properties down to $V\sim15$. Such a list is expected
to be useful for subsequent future studies of this sparse open cluster.

\subsection{Bona fide cluster members}

We used the following criteria to decide whether or not to assign 
the status of bona fide member to a probable cluster member:
\begin{enumerate}
\item Proper motion membership probability $P_{\mu}\geq5\%$, or all 
115 stars from Table~1.  This low $P_{\mu}$ threshold is
chosen in order to essentially preserve all possible cluster members.
\item Within the photometric errors a star is on the zero-age main sequence
(ZAMS) or above it due to the binarity and/or the pre-main sequence status
at $V>12$. 
\item The radial velocity is within $\sim$3 km~s$^{-1}$ around the cluster's
mean $<$RV$>=$$+$14.8 km~s$^{-1}$. Since most of our radial velocity
measurements 
were obtained at a single epoch, this limit works against the single-lined
binary cluster members that may have their instantaneous radial 
velocity outside the 3 km~s$^{-1}$ window. If a star is a known 
or suspected binary, this criterion is relaxed.
\item At the age of IC~2391 ($\sim$40 Myr), the stars cooler than 
$T_{\rm eff}\sim$ 7000~K must have a Li~I feature in their spectra at 6708\AA. 
The lack of lithium indicates that a star is a much older field star.
The Li test could not be applied to several fainter stars due to the
lack of high-resolution spectra.
\end{enumerate}

If any of the criteria listed above fails, such a star may not qualify for
a list of bona fide cluster members. In the case of double-lined spectroscopic
binaries their barycentric velocity should be within $\sim$5 km~s$^{-1}$
around the cluster's mean velocity.
Thus, a total of 66 bona fide cluster members
are selected and denoted by an asterisk in the last column
of Table~1.  For convenience, we also provide cross-identifications
with HD stars (numbers greater than 70000) in Table~1 and the
VXR numbers from \citet{pat96}.
It is still possible that a few
stars among these stars may not survive further scrutiny for 
cluster membership, and some rejected stars may turn out to be genuine
cluster members.

A list of bona fide cluster members was used to find the cluster center,
previously not very well known for IC~2391. We assumed a Gaussian profile
to model marginal distributions in right ascension and declination
of the observed star spatial density. 
A fit to these distributions yields a cluster center, equal to
$\alpha$=8$^{\rm h}41^{\rm m}00\fs3$ and $\delta$=$-53\degr00\arcmin36\arcsec$
(J2000.0). Due to a highly asymmetric distribution of cluster members,
the uncertainty in the cluster center is about $1\arcmin$.

\begin{figure}
\resizebox{\hsize}{!}{\includegraphics[angle=-90]{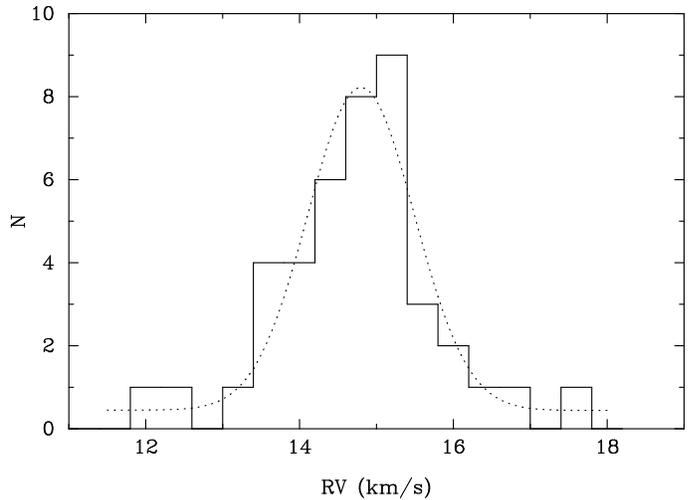}}
\caption{Histogram of mean radial velocities for bona fide cluster members
of IC~2391.  The dashed line shows a Gaussian fit that peaks at 
$+14.80$ km~s$^{-1}$. 
}
\label{fig:rv_hist}
\end{figure}

The sample of bona fide cluster members allowed us to re-examine the 
distribution of radial velocities drawn nearly exclusively from this
study only, i.e., the merged list of \textsc{Coravel} and
FEROS radial velocities. 
We selected 42 stars showing no apparent signs of duplicity. 
The distribution of radial velocities was binned in 0.4 km~s$^{-1}$
increments and then fitted with a Gaussian (Fig.~\ref{fig:rv_hist}). The best fit yields a mean
radial velocity of 14.80$\pm0.69$ km~s$^{-1}$ (s.d.). This is 
very close to the  mean radial velocity of
14.6 km~s$^{-1}$ estimated from the \citet{sta97} data.
The distribution of radial velocities appears to be slightly skewed.
We believe that small number statistics is a primary source of this
skewness, although stars with $v\sin i>90$ km~$^{-1}$ 
may also contribute to this effect by having their radial velocity slightly 
increased. 

A few stars in Tables~\ref{tab:meanrv} and \ref{tab:meanli} that have their
radial velocities very close to the cluster's mean velocity fail one or two
additional criteria. Two of them, 2717 and 5314, are
located below the main sequence but otherwise would qualify for
cluster membership and probably deserve further scrutiny. The third
star, 5376, is almost a perfect cluster member, notwithstanding the 
noted absence of lithium. The available spectrum is rather noisy;
therefore, we cannot be confident about the lack of Li feature, so
additional checks are required. According to \citet{ran01}, the
cluster member 4636=SHJM~9, only by $\sim$100 K hotter than star 5376, has
a fairly prominent Li~6708 feature with EW(Li)=100 m\AA. Hence,
there is no reason to assume that Li would be depleted below detection
in star 5376, if it were a cluster member.

\begin{figure*}
  \begin{center}
  \begin{tabular}{c}
  \resizebox{0.9\hsize}{!}{\includegraphics[angle=-90]{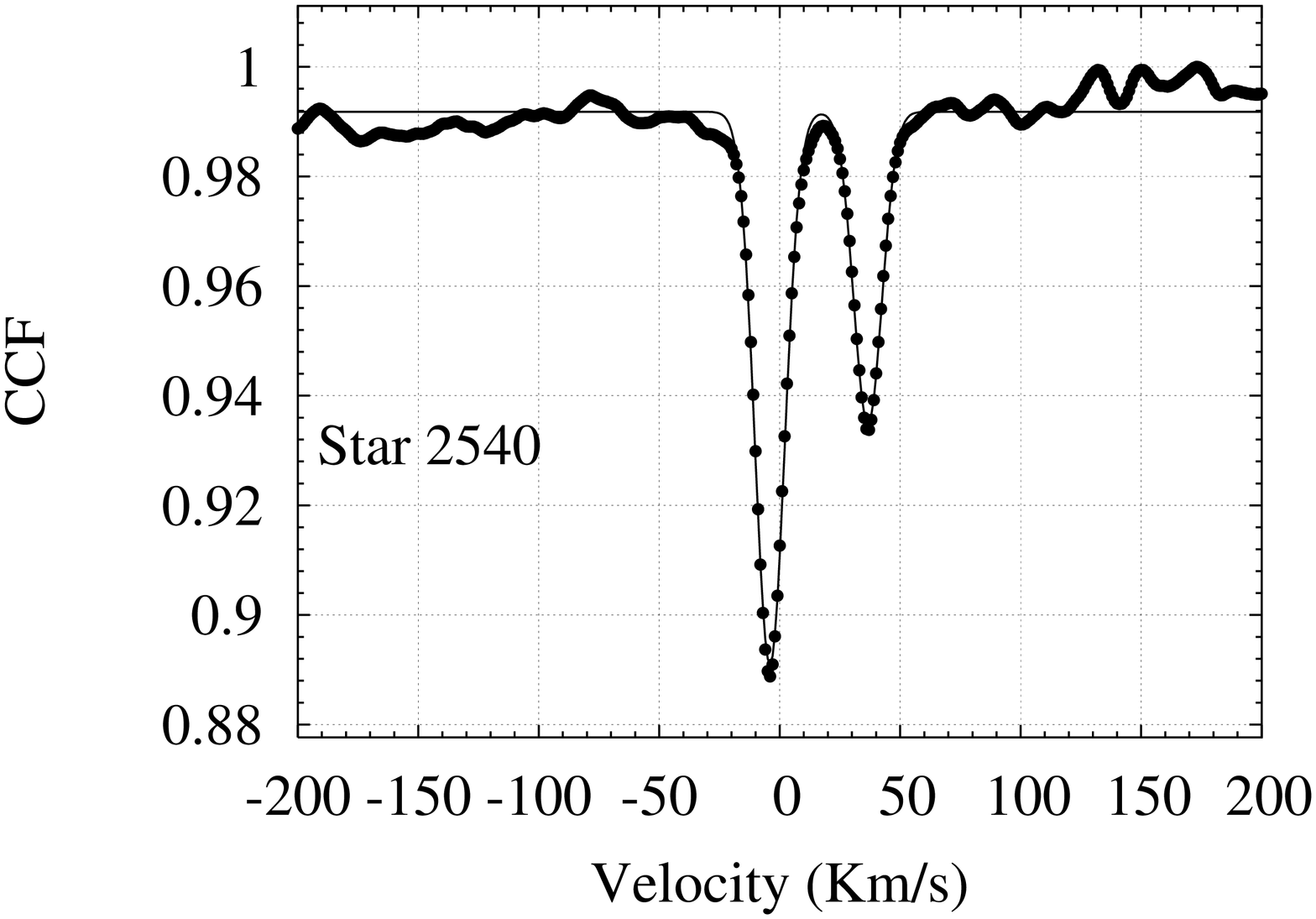}\includegraphics[angle=-90]{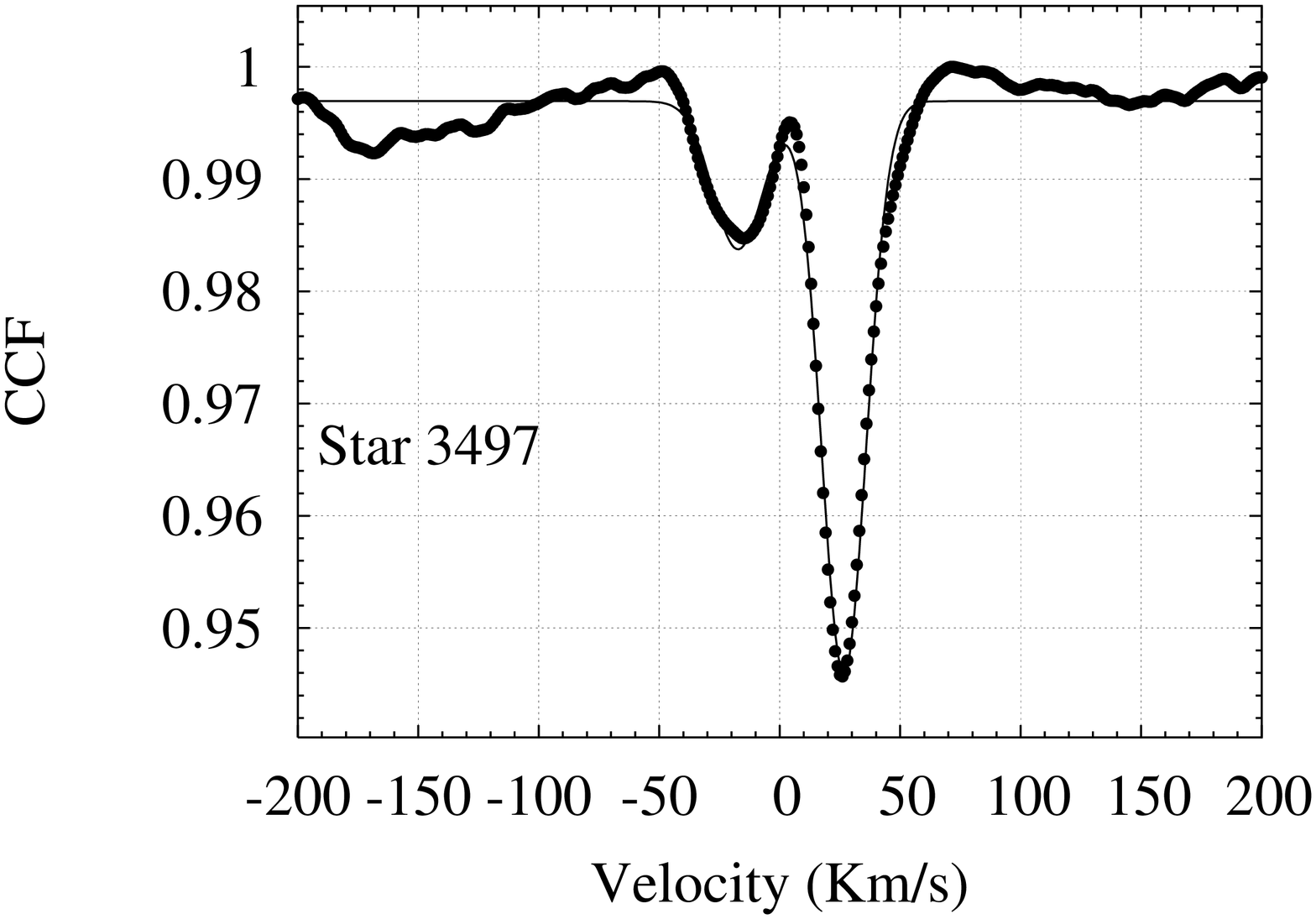}}\\
  \resizebox{0.9\hsize}{!}{\includegraphics[angle=-90]{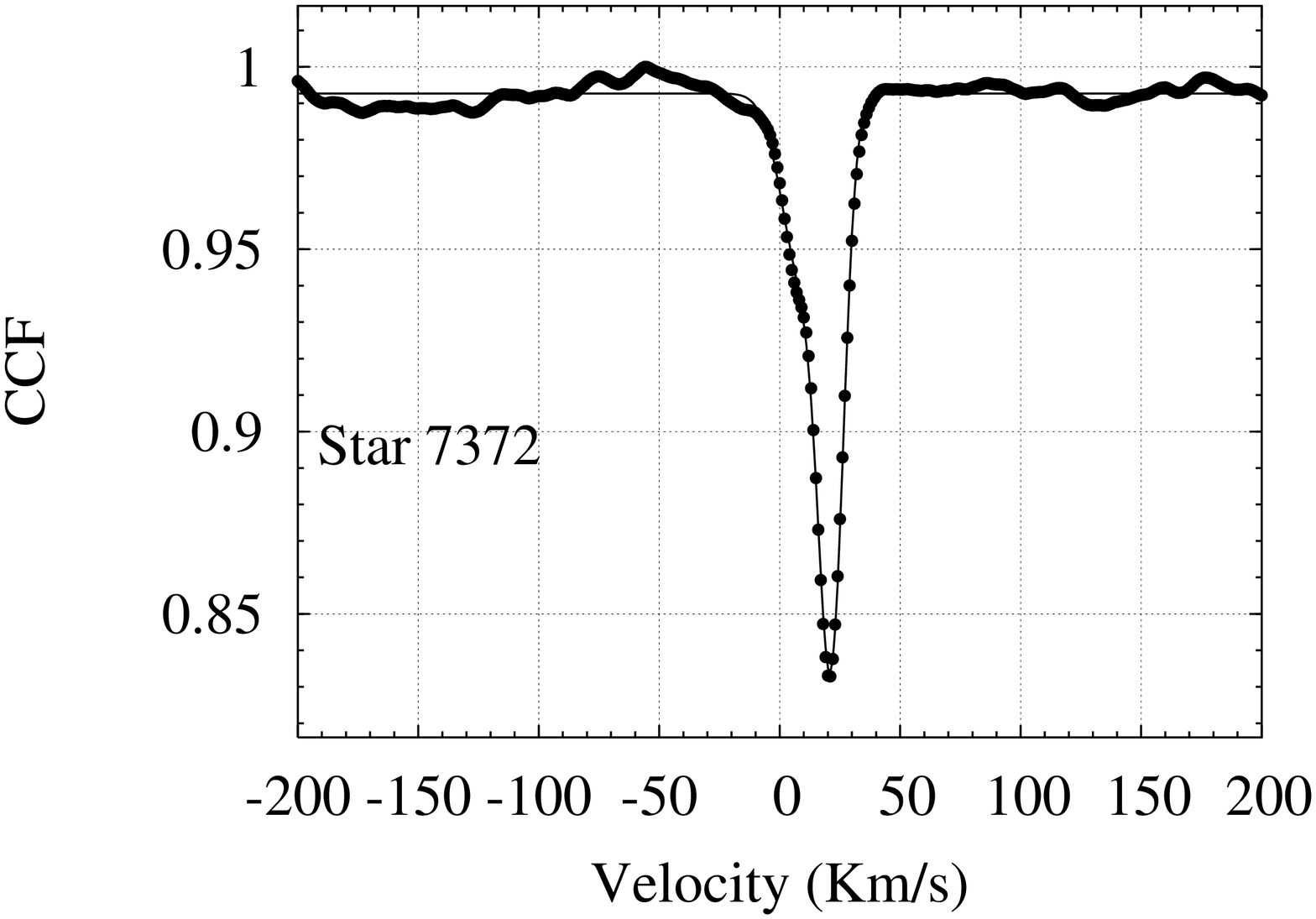}\includegraphics[angle=-90]{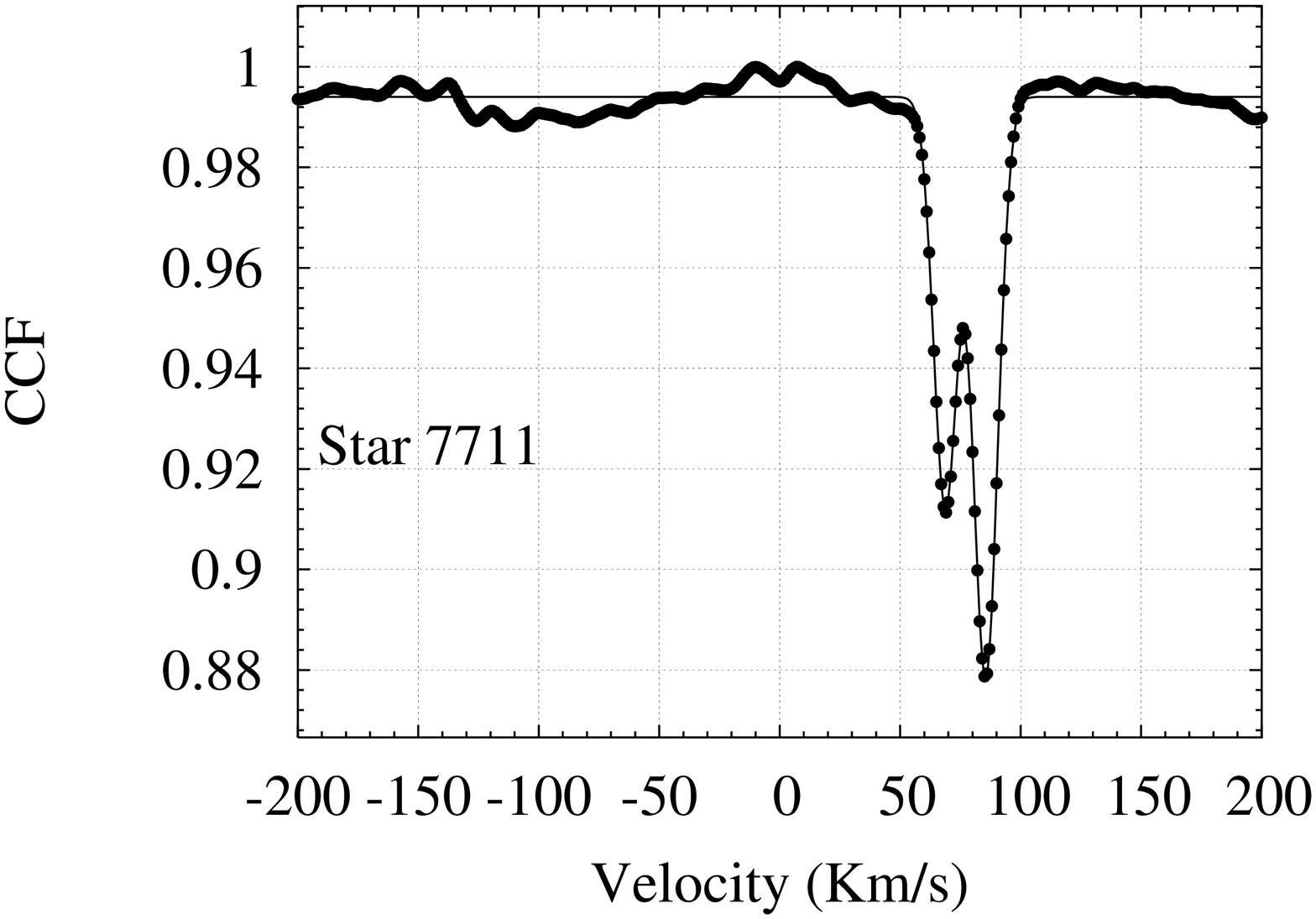}}\\
  \end{tabular}
    \end{center}
\caption{
Examples of cross-correlation function for double-lined spectroscopic 
binaries found among the FEROS target stars.
}
\label{fig:sbs}
\end{figure*}

\subsection{Membership status of Patten \& Simon and SHJM stars}

Many of the recent advances in the understanding of IC~2391 rest on the X-ray
source list from ROSAT observations \citep{pat93,pat96}. A total of
80 sources have been identified from these observations. Owing to a rather
low spatial resolution ($\sim$5-30$\arcsec$ FWHM) for ROSAT detectors,
identification of optical counterparts in many cases is uncertain.
To ensure that none of the possible counterparts with $V<19$ is missing, 
\citet{pat96} provide a total of 184 possible identifications 
and then try to narrow down the list using available proper motions,
photometry, and spectroscopy. In essence, the papers by \citet{sta97},
analyzing 26 counterparts, and \citet{ran01}, re-analyzing a
subsample from the Stauffer et al. paper, is a continuation of this 
effort. We have astrometric data for 49 optical counterparts,
all brighter than $V\sim15$
(see cross-identifications in Table~\ref{tab:patten}). Among these stars, 12 
are field stars with the membership probability, $P_{\mu}$=0.
A formal sum of probabilities indicates that the expected number of
cluster members is $\sim$25. This estimate is lower
than the actual number of members because we could not incorporate
radial distribution in the membership calculation (see Sect.~2).
Indeed, there are 34 bona fide cluster members among 
the likely optical counterparts of X-ray sources. We conclude that,
while proven successful in finding many low-mass cluster members,
the X-ray activity alone is not a decisive and comprehensive
membership criterion.

To illustrate how deceptive corroborative evidence can sometimes
be, consider star 6576=WXR 50a. It is listed as a ``suspected
cluster member based on photometry and/or spectroscopy'' by \citet{pat96}.
Its radial velocity from our study is +15.13  km~s$^{-1}$ and
+17 km~s$^{-1}$ \citep{sta97}, both very close to the cluster's
mean radial velocity. For this star, the Li abundance is 2.7 (our study)
and 2.6 according to \citet{ran01}, tightly following the
$\log$ N(Li) vs. $T_{\rm eff}$ trend for IC~2391 (Fig.~\ref{fig:li}). 
A closer inspection of color-magnitude diagrams
$V,(V-I)$ and $J,(J-K)$ shows that a star is located 
$\sim$0.5 mag {\it below} the main sequence.  However, it is 
the proper motion of 6576 ($\mu_x=-8.8$ and $\mu_y=+2.6$ mas yr$^{-1}$)
that, at $11\sigma_c$ apart from the cluster's mean motion, 
rules out the cluster membership. 

One star, 4658=VXR~45a, warrants a special statement. While it is not
listed by us as a bona fide member because it is located relatively far
above the main sequence in the color-magnitude diagram,
its high proper-motion membership probability of $P_{\mu}=75\%$
strongly suggests the cluster membership. 
This G9 spectral type star, also a BY Dra
type variable V370 Vel, stands out by its huge rotational velocity 
of $v \sin i =$238 km~s$^{-1}$ \citep{mars04}, by far exceeding 
any other $v \sin i$ measurement in IC~2391 
\citep[][ see also Tables~\ref{tab:meanrv} and \ref{tab:meanli}]{sta97}. The short photometric rotational period equal to
$P_{\rm rot}=0.223$ days \citep{pat96} is also clearly visible in the X-ray
time series observations \citep{mar03}.  If this is a genuine cluster
member, then the question is what mechanism spun it up or prevented the 
dissipation of its primordial angular momentum, while most cluster
members similar to star 4658 are slow rotators ($v \sin i<20$ km~s$^{-1}$).
It should be noted that such ultrafast rotators can be found in
other young open clusters, e.g., in the Pleiades \citep{vlee87},
$\alpha$~Per \citep{pro92}, IC 2602 \citep{sta97}. One possible
explanation of this phenomenon is offered by \citet{bars96},
invoking a paradigm of the angular momentum loss saturation.

There is another fast rotator VXR~80a with  $v \sin i \sim$150 km~s$^{-1}$ 
\citep{sta97} among the suspected
cluster members by \citet{pat96}. Since we have not measured this
star, the only source of proper motion for VXR~80a=King~391 is
a study by \citet{kin79}. In this paper the cluster's proper motion centroid is
at $\mu_{x}=-17.0$ and $\mu_{y}=+16.3$ mas yr$^{-1}$, derived by using
the counterparts of our bona fide cluster members in \citet{kin79}. 
If we adopt the standard error of proper
motions $\sigma$=1.5 mas yr$^{-1}$, then the proper motion of King 391 is
more than 6$\sigma$ apart from the cluster's mean motion which 
rules out the cluster membership. 

The star SHJM~5 also shows $v \sin i \sim$150 km~s$^{-1}$ \citep{sta89}.
This star is a member of a visual binary (our star 4757) for which we
have the membership probability $P_{\mu}=28\%$, consistent with cluster
membership. Finally, star 2457 with $v \sin i =$177 km~s$^{-1}$ is
another possible cluster member despite its measured radial velocity at
$-7.25$ km s$^{-1}$, since its astrometric membership probability is
$P_{\mu}=71\%$ and it is located just slightly above the main sequence in
both $BV$ and $JK$ color-magnitude diagrams.

Prior to ROSAT measurements, the so-called SHJM list of 10 relatively 
faint cluster members was published by \citet{sta89}. 
Cross-identifications of SHJM stars are provided by \citet{pat96}.
Based on our astrometry and criteria listed in Sect.~6.1, stars
SHJM 1,2,6,9,10 appear to be bona fide cluster members, while
SHJM~8 was not measured but SHJM~7=3664 has a membership probability of zero.
The membership for SHJM~3=1820 and (SHJM~4 + SHJM~5)=4757 is not
definitely constrained by our data. For SHJM~3 this is because
its radial velocity from our data is +12.76 km~s$^{-1}$, while 
\citet{sta89} provide +19.5  km~s$^{-1}$, thus
indicating a possible spectroscopic binary.
Its astrometric membership probability is only 9\%, which is more
a characteristic of a field star. 
Our conservative lower limit of equivalent width for the Li feature
for SHJM~3 is 2.5 times larger than the measurement by \citet{sta89}, thus
implying unusually high abundance of Li at $T_{\rm eff}$=4480 K
in the context of the overall $\log$ N(Li) vs. $T_{\rm eff}$ curve for IC~2391
(see Fig.~\ref{fig:li}). 

\subsection{Membership of some bright stars}

{\bf 3518=HIP 42504}: a 5th magnitude star whose Hipparcos astrometry
is consistent with cluster membership. Its redward position from 
the main sequence in the $BV$ and $JK$ CMDs (see Sect.~6.5) indicates 
duplicity, which is confirmed by radial velocity measurements \citep{fei61,lev88}. The
star is an SB1 spectroscopic binary with P=3.2d, eccentricity
$e=0.05$, and $\gamma$-velocity$=14.5$ km~s$^{-1}$. 

{\bf 3658=HD 74438}: this star is $\sim$0.9 magnitudes above the main sequence
in the color-magnitude diagram, indicating a potential triple system. 
Its proper-motion membership probability $P_{\mu}=88$\% is high
and reliable. \citet{fei61} lists two measurements of radial velocity:
$+7.5$ and $+17.5$ km~s$^{-1}$, whereas \citet{bus65} derives 
$+$21: km~s$^{-1}$ from five spectrograms. Thus, the scatter of radial
velocities is indicative of a non-single status of the star.
A formal mean velocity matches the cluster's radial velocity. We note
that at the XMM-Newton observatory \citet{mar05} were able to obtain 
only the upper limit of X-ray luminosity for this star, which is typical 
of A spectral type stars. On the grounds of the available data, we
consider star 3658 a bona fide cluster member, albeit one requiring
more spectroscopic studies.

{\bf 4484=HIP 42536=$o$~Velorum}: 
the brightest star in IC~2391, also a $\beta$ CMa
type variable. Its Hipparcos parallax $\pi=6.59\pm0.51$ mas is very close
to the mean parallax of IC~2391 equal to 6.85$\pm$0.22 \citep{rob99};
however, Hipparcos proper motion in declination 
$\mu_{\delta}=+35.1\pm0.5$ mas~yr$^{-1}$ is 20$\sigma$ (!) apart from
the cluster's mean proper motion $\mu_{\delta}=+22.7\pm0.2$ \citep{rob99}. 
\citet{gon01} report a significantly lower absolute proper motion 
$\mu_{\delta}=+23.6\pm2.0$ consistent with \citet{rob99}. Our
membership probability, $P_{\mu}=81$\%, and the mean radial velocity of
$+15.2$ km~s$^{-1}$ by \citet{vhoo72} together indicate that the star
is a bona fide cluster member. The discrepant Hipparcos proper 
motion can be explained as the effect of unresolved binarity \citep{wie99}.

{\bf 5459=HD 74009}: a star with lower membership probability ($P_{\mu}=25\%$)
than the other relatively bright stars. However, the mean radial velocity
of $+14$ km s$^{-1}$ by \citet{lev88} strongly supports cluster
membership. Apparent variability of the observed 
radial velocity \citep{lev88} indicates a possible spectroscopic binary.

{\bf 7847=HIP 42823}: another bright star in IC~2391 first considered to
be a member by \citet{egg91}. Its Hipparcos parallax is only $0.7\sigma$
apart from the mean parallax of IC~2391, while Hipparcos proper motion 
deviates from the mean by $\sim$3 mas~yr$^{-1}$ in both
coordinates. Our membership probability of 78\% supports the 
association of star 7847 with IC~2391, however, the lack of radial velocity
measurements prevents us from assigning it the status of a definite cluster
member.

\subsection{Spectroscopic binaries}

Identifying spectroscopic binaries in star clusters has a dual purpose.
First, the evolutionary paths of stars are reasonably well-understood only
for single stars; therefore, it is critical that we can identify these
stars in the color-magnitude diagram with high confidence. Radial velocities
from high-resolution spectroscopy is a powerful tool for detecting close 
binaries in a wide range of mass ratios. Second, the binaries themselves 
are very important to astronomy, especially those in the star clusters 
that provide a coeval sample of stars in a wide range of masses, all
having the same initial composition.

The early spectroscopic work in IC~2391 was focussed on bright, early
type stars \citep{fei61,bus65,per69a,vhoo72,lev88}. In Table~1
-- last column -- we list only double-lined spectroscopic binaries (SB2),  
single-lined spectroscopic binaries (SB1 or SB as listed in the source),
and suspected 
spectroscopic binaries (SB:), if a star shows a variable radial velocity
or the mean radial velocity in different studies is significantly
variant. The binaries with orbital solutions are marked with a 
letter `o'. Since the internal mean error in these studies can reach
up to 3-5 km~s$^{-1}$, not all cases of suspected SB may be real.

\begin{figure}
  \resizebox{\hsize}{!}{\includegraphics[angle=0]{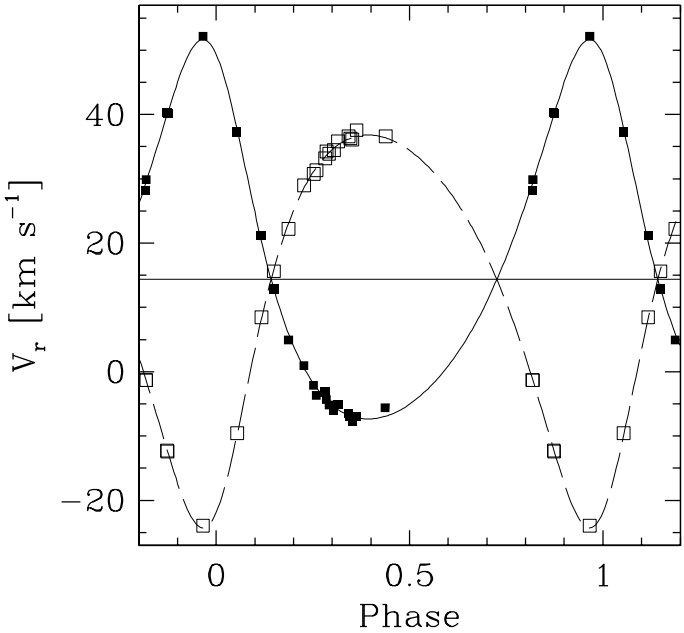}}
\caption{Radial-velocity curve for spectroscopic binary star 4413 convolved 
with the orbital period of P=90.617 days. The filled squares denote component
A; open squares -- component B. 
}
\label{fig:orb}
\end{figure}

Among the IC~2391 F-K type main sequence stars, the long-term \textsc{Coravel} 
observations have revealed two SB2 and four SB1 spectroscopic binaries 
(see Tables~3 and \ref{tab:meanrv}). An SB2 spectroscopic binary
4413 has a sufficient number of observations
to calculate the orbital elements listed in Table~\ref{tab:param}. This binary
system (Fig.~\ref{fig:orb}) consists of two nearly equal-mass components 
on a fairly eccentric orbit ($e\sim0.3$). Its $\gamma$-velocity of
14.35 km~s$^{-1}$ clearly supports the cluster membership.

According to its proper motion, $\mu_x=-4.0$ and $\mu_y=+11.8$ mas yr$^{-1}$,
the SB2
spectroscopic binary 389 is a field star. As indicated by \textsc{Coravel}
observations in Table~3, this binary was not
resolved over four epochs, and these observations yield an estimate
of the $\gamma$-velocity at 9.27 km~s$^{-1}$ (Table~\ref{tab:meanrv}). 
The \citet{wil41} method allows us to determine the $\gamma$-velocity more
precisely and to obtain the mass ratio in this binary system. Thus,
the refined $\gamma$-velocity is $+10.24$ km~s$^{-1}$ 
(K$_{1}=19.36$ and K$_{2}=21.75$  km~s$^{-1}$) and the mass ratio 
$M_a/M_b=0.89$. Clearly, the $\gamma$-velocity of star 389 also rules out
cluster membership.

Finally, the FEROS high-resolution spectra obtained in one or two epochs
(Table~5) indicate five SB2 binaries, four of them 
(819, 2540, 3497, 7372) are bona fide cluster members. For a set of four
FEROS double-lined spectroscopic binaries, the observed cross-correlation
functions are shown in Fig.~\ref{fig:sbs}. One of them (star 7711)
has a barycentric velocity clearly incompatible with the mean radial
velocity of the cluster. 

Altogether, among 66 bona fide members, there are seven SB2, six SB1, and 
nine suspected spectroscopic binaries. If all spectroscopic binary
categories are considered, then the binary frequency in IC~2391 is 
at least $\sim$30\%. The location of all 22 spectroscopic binaries in $BV$ 
color-magnitude diagram is shown in Fig.~\ref{fig:cmd_zams}.

\subsection{Color-Magnitude Diagram}

\begin{figure}
  \resizebox{\hsize}{!}{\includegraphics[angle=0]{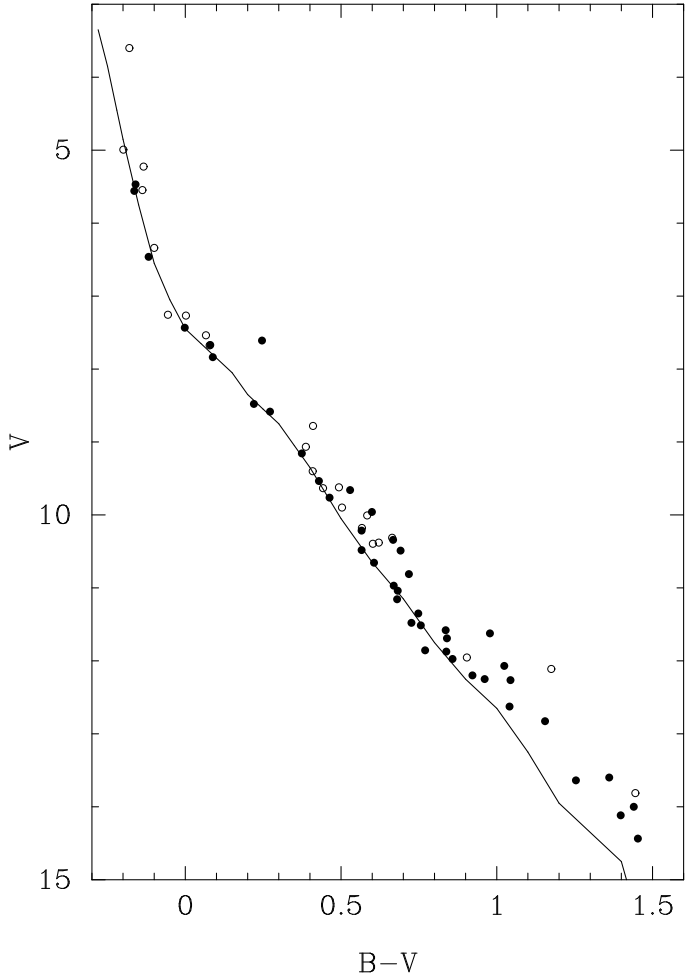}}
\caption{Color-magnitude diagram for the bona fide members of
IC~2391. The bold dots denote single stars or the stars with unknown
multiplity status. The open circles indicate spectroscopic binaries.
Zero-age main sequence is overplotted using the distance modulus
equal to 5.95 mag and zero reddening, $E(B-V)$=0.0.
}
\label{fig:cmd_zams}
\end{figure}

There are two sources of complete photometry for the bona fide cluster
members: our CCD $BV$ and 2MASS $J\!H\!K_s$ \citep{skr06} photometry. 
In the case of a missing CCD $BV$ photometry value, we used
photoelectric $BV$ photometry from the literature. Such values in
Table~1 are recognizeable by having only two decimal digits. 
The best previous CMD in $U\!BV$ is presented 
by \citet{per69b}. These authors conclude that IC~2391 is unreddened, i.e,
$E(B-V)=0.00$.  It was later confirmed by measurements in the 
Vilnius seven-color photometric system \citep{for98} and is also adopted 
in the present paper. The true reddening probably is not exactly zero;
however, the limited precision of existing $U\!BV$ photometry and still unclear
binarity status of some early spectral type cluster members preclude us
from deriving a better reddening estimate.
The following fits to the CMD, however, do not support
the high value of $E(B-V)=0.06$ reported by \citet{bar04}. 
The studies by \citet{sta89,sta97} and \citet{pat96} 
mainly employ $V\!I$ photometry in constructing the CMD of IC~2391,
assuming a distance modulus (m-M)=6.05.
We note that the latter is 0.23 mag larger than the distance
modulus from Hipparcos parallaxes \citep{rob99}. The color-magnitude
diagrams in these studies indicate that the stars at $V\geq11$ are
located above the main sequence, presumably still being on the
pre-main-sequence tracks. Our list of bona fide cluster members
allows us to explore the properties of the $BV$ CMD (Fig.~\ref{fig:cmd_zams})
in more detail.

We made a trial fit  of ZAMS \citep{lan82}
to the $BV$ CMD, assuming zero reddening, $E(B-V)=0.0$, and a distance
 modulus of
$V_{0}-M_{V}$=5.95. It appears that bright stars are essentially
unevolved, with the exception of the brightest member, 
star 4484 = $o$~Velorum, whose color is suspect due to possible 
duplicity (see Sect. 6.3). The lower main sequence seems to end at
$V\ga12$ giving rise to pre-main sequence stars at fainter
magnitudes. An exact interpretation of the $BV$ color-magnitude
diagram for young open cluster clusters is complicated by the fact
that K~dwarfs in the Pleiades are either subluminous and/or have
abnormally blue $B-V$ color \citep{sta03}, hence falling below
the ZAMS. These authors suggest that all young K~dwarfs may
show a similar anomaly, thereby limiting the ability to obtain
an estimate of the so-called PMS isochrone age. We examine this effect
in the following section.

\subsubsection{Main-sequence fittting, distance modulus, age}

One of the goals of this study is to obtain a reliable photometric distance
from the main sequence fitting. The empirical ZAMS
used in Fig.~\ref{fig:cmd_zams} is not adequate for at least two
reasons. First, the ZAMS from \citet{lan82} is based on the Hyades distance
modulus of $m-M=3.28$ mag , while Hipparcos parallaxes for the Hyades members
yield 3.33 mag. Second, at the time of compiling this ZAMS, the
metallicities of open clusters were poorly known. The
uncertainty in [Fe/H] is directly related to the uncertainty of the
amount of
the metal-line blanketing effect on colors and magnitudes -- a major
source of systematic errors in photometric distances.
 
Recently, \citet{pin03,pin04} and \citet{an06} derived new isochrones
specifically targeted to measuring the distances to open clusters and
other parameters
from the main-sequence fitting. They used the Yale Rotating Evolution
Code (YREC) to construct stellar evolution tracks for masses
$0.2\le~M_{\sun}\le8$ and metallicities $-0.3\le[{\rm Fe/H}]\le+0.2$.
The tracks were interpolated to provide theoretical isochrones
for stellar ages from 20 Myr to 4 Gyr, with the pre-main sequence phase
included where appropriate. Then, the authors used
the empirical $T_{\rm eff}$-color transformations from \citet{lej98} and
applied small additional corrections to isochrones in order to match
the photometry for the Hyades, adopting its metallicity at [Fe/H]=+0.13 dex.

\begin{figure}
  \resizebox{\hsize}{!}{\includegraphics[angle=0]{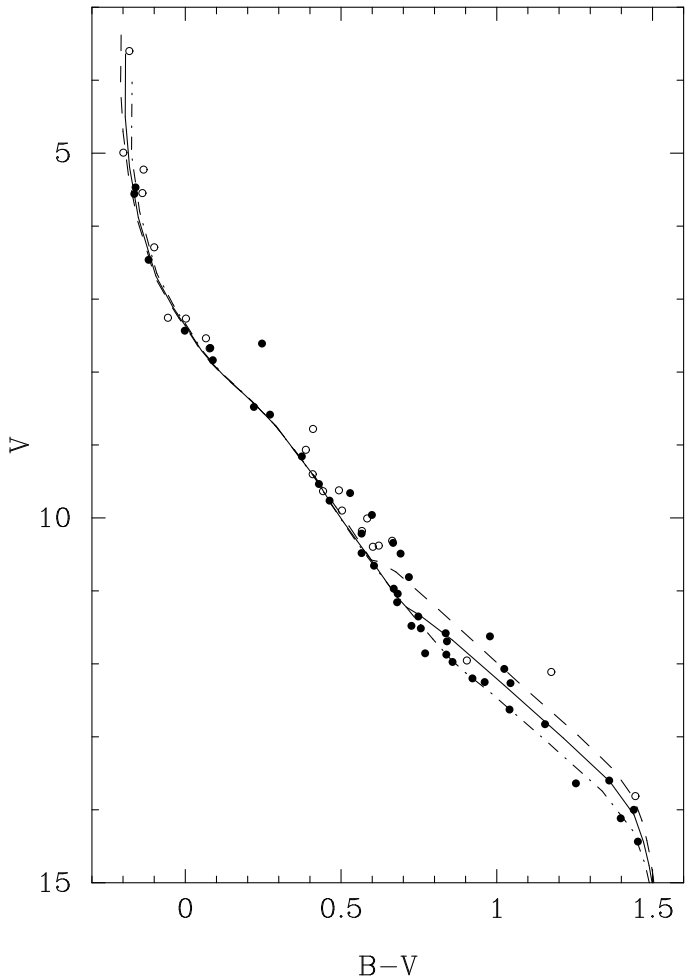}}
\caption{Color-magnitude diagram for the bona fide members of
IC~2391. 
See Fig.~\ref{fig:cmd_zams} for the meaning of bold dots and open circles.
The fit of 30, 40 (dark curve), and 55~Myr isochrones to CMD assuming
zero reddening ($E(B-V)$=0.0) yields the distance modulus of
$V_{0}-M_{V}$=6.01 mag. Our estimate of age is 40~Myr. The location
of stars below the 40~Myr isochrone at $V>11.5$ can be explained by
anomalous spectral energy distributions in young K dwarfs.
}
\label{fig:cmd_iso}
\end{figure}

We used these isochrones to derive a precise distance modulus and to
probe the range of possible cluster ages. The key parameter for assuring
reliability of main-sequence fitting is metallicity. Our new
spectroscopic estimate of [Fe/H] for IC~2391 has an accuracy similar to
the metallicities used by \citet{an06} in their analysis of four
open clusters. In previous studies the isochrone age of IC~2391 is
estimated to be $\sim$30-35 Myr \citep{sta97,mer81}, while the
location of the lithium-depletion boundary indicates an age of
$50\pm5$ Myr \citep{bar04}. We chose 30, 40, and 55 Myr isochrones
from \citet{an06} and interpolated them to match the cluster's
metallicity of [Fe/H]=+0.06 dex. These isochrones were fitted
to the $BV$ CMD (Fig.~\ref{fig:cmd_iso}) by adjusting them to the lower envelope
of main sequence in the color range of $0.2<B-V<0.7$ and by assuming
zero reddening, $E(B-V)=0.00$ mag. The resulting distance modulus
is $V_0-M_V=6.01$ mag, which is larger by  0.19 mag or $2.7\sigma$
than the distance modulus inferred from the mean parallax via 
the Hipparcos combined abscissae solution by \citet{rob99} and \citet{vlee99}.
Following the argumentation by \citet{pin98} and the study
of NGC~2451A \citep{pla01}, we adopted the uncertainty in our
distance modulus to be 0.05 mag.

This result should be considered in the context of a trigonometric
distance to the Pleiades from Hipparcos measurements vs. the Pleiades
distance estimate from the main-sequence fitting.
The Hipparcos measurements yielded a significantly smaller
distance modulus for the Pleiades (i.e., $m-M=5.37\pm0.07$ mag,
equivalent to $118.3\pm3.5$ pc in van Leeuwen 1999) than the distance
modulus from the main-sequence fitting, e.g., $m-M=5.60\pm0.04$
\citep{pin98}. 

It is distressing to find another open cluster whose Hipparcos
parallax distance is significantly shorter than the photometric
distance. We point out that among the eleven IC~2391 members used by
\citet{rob99}
to derive the mean cluster parallax, only one star (7027=HIP 42450)
has a smaller Hipparcos parallax than our photometric
parallax. Incidently, this star is the second faintest in the sample.
What kind of systematic errors could possibly bias our
photometric distance? The first is a small uncertainty in the reddening
of IC~2391. If the true reddening is as high as $E(B-V)=0.01$ mag,
that alone would reduce the distance modulus by 0.03 mag.
Second, if the true metallicity is solar, i.e., [Fe/H]=0.0 dex,
it would also reduce the distance modulus by an additional $\sim$0.08 mag.
However, it is unlikely (although not imposssible) that either
parameter is off by this much. Also, a major change in the stellar parameters
would be required to lower the [Fe/H] value enough to account for the
0.19 mag discrepancy in the distance modulus. 
 
In the light of another looming Hipparcos distance problem, it
is instructive to have a look of how it is resolved for the Pleiades open
cluster.  The following diverse and independent studies
actually all support the long distance to the Pleiades. Thus, \citet{gat00}
measured trigonometric parallaxes of seven members of the Pleiades
clusters and obtained an equivalent to the distance modulus
$m-M=5.59\pm0.12$ mag. Using the Hubble Space Telescope's Fine
Guidance Sensor, \citet{sod05} obtained absolute trigonometric parallaxes
for three Pleiades members leading to the distance modulus
$m-M=5.63\pm0.02$ mag. 
When \citet{pan04} applied Kepler's third law and the mass-luminosity
relation to their interfermetric measurements, they found that the Pleiades
visual binary star Atlas has $m-M=5.66\pm0.04$ mag.
A different set of interferometric observations
for Atlas, combined with the measurements of radial velocities,
produced better-constrained orbital elements and, consequently,
a purely geometric distance or $m-M=5.60\pm0.07$ mag
\citep{zwa04}.  More strong evidence comes from the combined
orbital solution of the eclipsing binary HD~23642, yielding the
distance modulus $m-M=5.60\pm0.03$ mag \citep{mun04}. Finally,
the frequency observations in six $\delta$ Scuti stars in the
Pleiades match the eigenfrequencies of rotating
stellar models best, when assuming $m-M=5.60-5.70$ \citep{fox06} and using
the other generally adopted parameters of the Pleiades. 

Regarding Hipparcos parallaxes, 
\citet{vlee05a} indicates that correlations in the abscissa residuals
for bright stars and weight disparities between each two fields-of-view
in the attitude reconstruction process can be attributed to the problematic
parallaxes for some open clusters, particularly the Pleiades.
\citet{mak02} has shown that by accounting for the average residuals
around the so-called reference circle helps to reduce the Pleiades
Hipparcos parallax to an equivalent of $m-M=5.55\pm0.06$. This,
however, is not the final word since the raw Hipparcos data are
under complete re-reduction \citep{vlee05b}.

Our $BV$ color-magnitude diagram (Fig.~\ref{fig:cmd_iso}) 
conclusively shows that the main
sequence extends as deep as $V$$\sim$12 or $M_v=+6$ and appears to
populate the pre-main-sequence tracks at fainter magnitudes.
If we adopt an age of IC~2391 equal to 40 Myr, our 40-Myr-old
pre-main sequence isochrone (see Fig.~\ref{fig:cmd_iso})
is located above the bulk of presumably pre-main-sequence stars.
That could be explained by the observed `blueshift' of K dwarfs
in the $BV$ CMD of Pleiades \citep{sta03}, noting that this effect
in IC~2391 starts at $B-V\sim0.7$ or at a spectral type G5.
A 55-Myr old isochrone significantly mitigates this effect but,
on the other hand, it does not fit the upper main sequence well,
suggesting that the isochrone (nuclear) age is less than 55 Myr.
In fact, the age estimate is constrained
just by the single brightest cluster member 4484 = $o$~Velorum. 
Because of its suspected duplicity 
(see Sect.~6.3) and a $\beta$CMa type variability \citep{vhoo72}, 
this star is not a typical B3 spectral type subgiant -- its color
$B-V=-0.18$ could be biased and, subsequently, affect the isochrone age.
Summarizing, the age estimate from the $BV$ CMD could be anywhere between
30 and 50 Myr.

\begin{figure}
  \resizebox{\hsize}{!}{\includegraphics{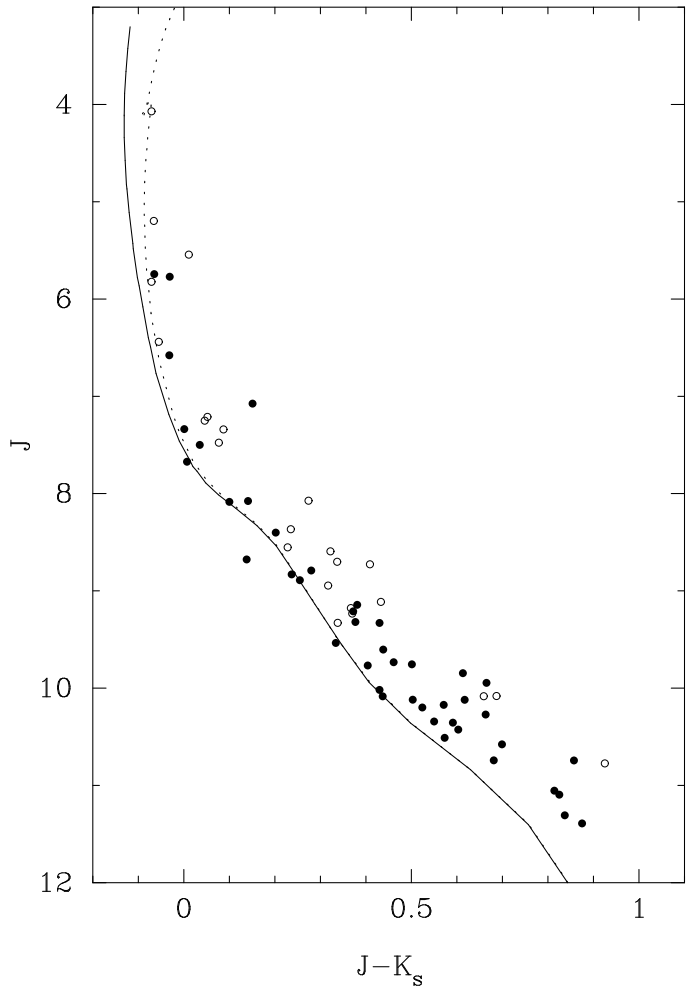}}
\caption{Color-magnitude diagram for bona fide members of IC~2391 using 
2MASS $J\!K$ photometry. 
See Fig.~\ref{fig:cmd_zams} for the meaning of bold dots and open circles.
The reddening and distance modulus are the same as in
Fig.~\ref{fig:cmd_zams}. Two 40 and 100~Myr solar metallicity isochrones
from \citet{bon04} are overplotted. 
}
\label{fig:cmd_2mass}
\end{figure}

Recently \citet{dod04} attempted to use the 2MASS photometry to construct
a CMD of IC~2391. One lesson learned from this exercise is that
the accuracy of 2MASS photometry is inadequate at $H>14$ ($J>15$).
Our bona fide cluster members are not fainter than $J=11.5$, so they
can serve as building blocks in constructing an empirical ZAMS from the
2MASS data. The $JK$ color-magnitude diagram of IC~2391 
(Fig.~\ref{fig:cmd_2mass}) has a considerably larger scatter across 
the main sequence than does the $BV$ CMD (Fig.~\ref{fig:cmd_iso}),
possibly due to the higher sensitivity to a binary presence. 
It is quite possible that several undetected spectroscopic binaries
are still hidden in our sample of bona fide cluster members. 
Some other reasons for this scatter may include
source confusion and different photometric measurement reductions
for stars brighter and fainter than $J\sim9$ \citep{cut03}.

In the $J\!K$ CMD (Fig.~\ref{fig:cmd_2mass}), we overplotted
40 and 100~Myr solar metallicity isochrones from
\citet{bon04} using the same distance modulus as for the initial ZAMS
fit in Fig.~\ref{fig:cmd_zams}, i.e., 
$V_{0}-M_{V}$=5.95. Only a handful of stars are located right on
the 40~Myr isochrone, which itself poorly fits the upper main sequence.
The best fit with a 100~Myr isochrone is
most likely compromised by some bias in the 2MASS photometry at bright
magnitudes and/or by using a $T_{\rm eff}$-color transformation that is
somewhat uncertain in this magnitude range.
Thus, choosing a 100~Myr isochrone helps to eliminate the mismatch at
the bright end of main sequence; however, it is very unlikely that IC~2391
is indeed that old.

\begin{figure}
  \resizebox{\hsize}{!}{\includegraphics{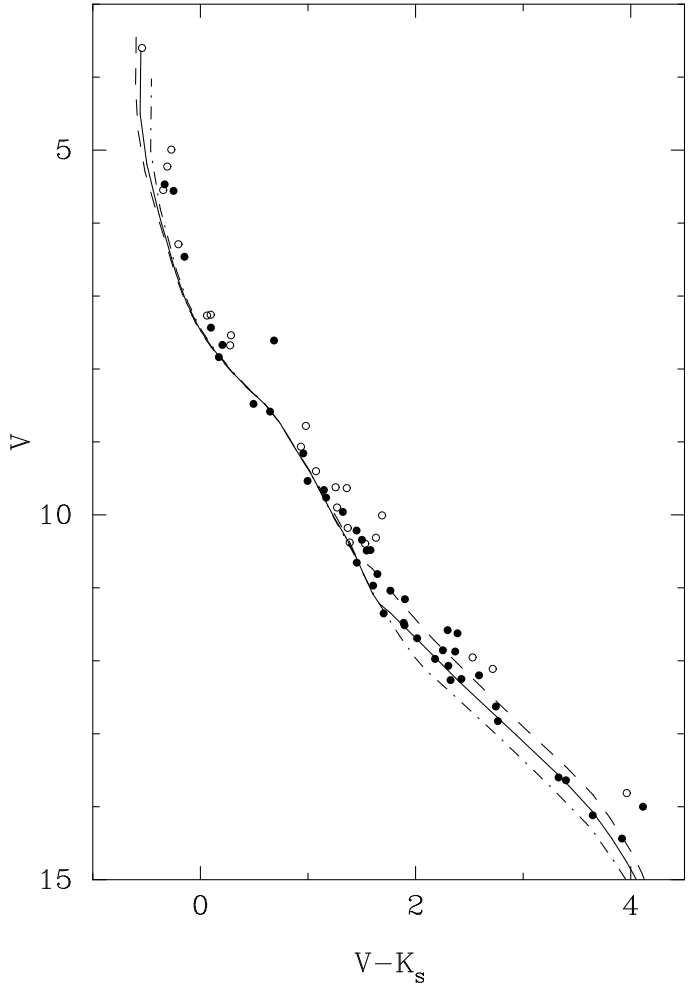}}
\caption{$V\!K$ color-magnitude diagram for bona fide members of IC~2391.
See Fig.~\ref{fig:cmd_zams} for the meaning of bold dots and open circles.
The reddening and distance modulus are the same as for
Fig.~\ref{fig:cmd_iso}. The best fit to the lower part of CMD is provided
by a 40~Myr isochrone (dark curve). For comparison, 30 and 55 Myr 
isochrones
are also plotted. Similar to $J\!K$ CMD, a fit to the upper part of
main sequence is poor most likely for the same reasons as for the CMD
from 2MASS photometry (see Fig.~\ref{fig:cmd_2mass}).
}
\label{fig:cmd_vk_iso}
\end{figure}

Following the referee's suggestion we investigated the $V\!K$
color-magnitude diagram.  In these colors it is expected that the K dwarf
`blueshift' in the CMD would disappear \citep{sta03}. Our $V\!K$ CMD
for bona fide cluster members is given in Fig.~\ref{fig:cmd_vk_iso}.
We attempted to fit this CMD with 30,35,40,45,50,55-Myr isochrones; however,
a relatively poor match to the essential part of main sequence
comprising F-G spectral type stars ($0.3\le~V-K_s\le1.5$) prevented
us from improving the distance modulus over a similar fit of the $BV$ CMD. 
Hence, we adopted the distance modulus of $V_0-M_V=6.01$ obtained
from the $BV$ CMD.  It is difficult to judge what causes
the scatter across the main sequence and the offset of its upper
portion. Possibly, the 2MASS photometry contributes to these
effects, but we also cannot entirely rule out small isochrone calibration
problems for the $V-K_s$ color index. On the other hand, the lower part
of color-magnitude diagram is indeed well-defined and allows us to confirm the
best estimate of isochrone age for IC~2391 at 40 Myr.

The so-called lithium-depletion boundary (LDB)
age method yields a higher age for IC~2391
at 50-53 Myr \citep{bar99,bar04} and 50 Myr by \citet{jef05} using
five different models. It should be noted that this is in line with similarly
higher ages by this method for the $\alpha$ Per cluster \citep{sta99} and
the Pleiades \citep{sta98}. Since our data on IC~2391 do not reach the
lithium-depletion boundary, we are not in a position to resolve
the age differences. 

\subsection{Rotational velocities $v\sin i$}

\begin{figure}
  \resizebox{\hsize}{!}{\includegraphics[angle=-90]{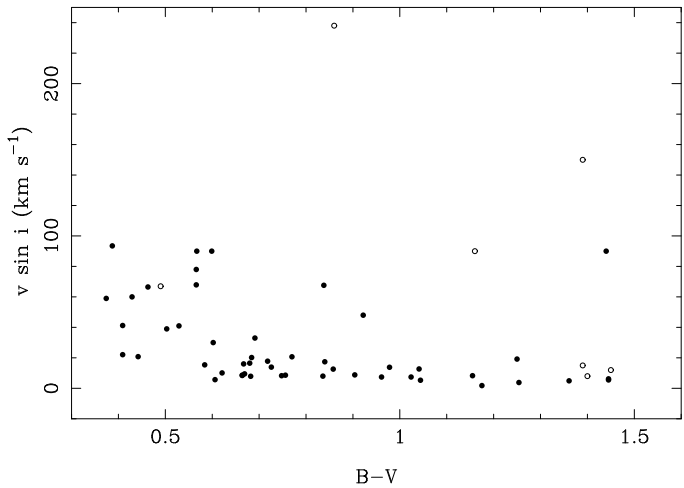}}
\caption{Distribution of $v \sin i$ for bona fide cluster members
as a function of $B-V$, which is roughly inversely proportional to luminosity.  
The filled circles are new measurements; open circles are those 
from the literature (fast rotators 1820, 4658, 4757 -- likely but
not bona fide cluster members -- are also included).
The distribution is bimodal, peaking at 10 and $\sim$80 km~s$^{-1}$.
See Sect. 6.2 about the ultrafast rotator, star 4658.
}
\label{fig:vsini}
\end{figure}

The distribution of $v\sin i$ for cluster members ranges from $\sim$2
to 240 km~s$^{-1}$ (Fig.~\ref{fig:vsini}).  Such a spread has already
been observed by \cite{sta97} for the IC~2391+IC~2602 members and can be 
interpreted as a result of the early angular momentum
evolution of low-mass stars, which appears to be regulated by the 
disk-locking mechanism \citep[e.g.,][]{edw93}. 
According to this scenario, a gradual dissipation of the disk weakens 
the magnetic coupling between the star and its circumstellar disk, thereby
releasing the star to spin up as it contracts during its pre-main-sequence
phase.

\cite{her05} analyzed a large data set of rotational periods for 
500 low-mass stars in five nearby young open clusters (Orion nebula cluster, 
NGC~2264, $\alpha$~Per, IC~2602, and the Pleiades). They show that 50-60\% 
of the stars still on the convective tracks, i.e., the vertical part
of the pre-main-sequence tracks, 
appear to be released from the locking mechanism early
and thus account for the fast rotators. 
Conversely, the remaining stars lose a considerable amount of their
angular momentum in the first few million years and enter onto the ZAMS
as slow rotators.
The distribution of $v\sin i$ in IC~2391 (Fig.~\ref{fig:vsini}) seemingly
follows this pattern. However, a relatively short empirical
life-time of circumstellar disks of $\sim$6 Myr \citep{hai01} makes
IC~2391, at its age of 40 Myr, an improper environment for testing the 
disk-locking mechanism. Recently, \citet{bar03} put forward 
a new interpretation of the observed pattern of rotational velocities
(periods) in star clusters like IC~2391 and older. In this interpretive
paradigm, the observed rotational morphology in G, K, and M stars
along the so-called $I$ and $C$ sequences is driven by dynamos of two
types that evolve in synchrony with the extent of the convective zone in stars.
For IC~2391 the only source of rotational periods for 16 stars is
the study by \citet{pat96}. We hope that the much larger sample of $v\sin i$
obtained in this study will help to advance the understanding of
rotational evolution.

\section{Conclusions}

A decade ago \citet{sta97} commented that IC~2391, as well as IC~2602,
are not giving up their secrets easily. Placed at an optimum age, 
these two clusters are ideal laboratories for studies of interface
between the main sequence and late pre-main-sequence phases and their
timescales. However, the intrinsic paucity of cluster members
in both IC clusters is the main factor limiting the scientific
return in the sense that even a single peculiar cluster member or
an interloper may distort the observable trends and relationships. 
One such sensitive case is star 1820 (see Sect. 6.2) and its location
on the Li abundance curve.

In this study we have significantly advanced the knowledge of IC~2391
in the range $-2<M_v<+8.5$ or down to $\sim$0.5M$_{\sun}$. This is
a first extensive proper-motion study in the region of IC~2391 providing 
membership probabilities over a 9 deg$^2$ area, in effect assuring 
a high-degree of completeness in the chosen ranges of magnitudes,
as explained in Sect.~2.  We have measured the radial velocity for most
of them. The new kinematic members
mainly fill in the F2-K5 spectral region. The \textsc{Coravel}
measurements of 31 stars considerably strengthen the membership
status for most of them. Several spectroscopic binaries were detected;
among them seven are double-lined SB2, with one yielding an orbital
solution.  The FEROS high-dispersion spectra served to measure
radial velocities and projected rotational velocities and to
estimate Li and Fe abundance. The latter is found to be
[Fe/H]$=+0.06$ on the scale of solar abundance 
$\log \epsilon$(Fe)=7.45 dex. 

A total of 66 bona fide cluster
members are selected by combining kinematic, spectroscopic, and 
photometric membership criteria. The $BV$, $J\!K$, and $V\!K$ 
color-magnitude diagrams were constructed using the bona fide
cluster members. The main sequence
fit yields a distance modulus of $V_{0}-M_{V}$=6.01 mag or 159.2~pc
that is significantly larger than the distance modulus and distance
(5.82 mag and 146~pc) from the Hipparcos mean parallax for IC~2391.

The problem with Hipparcos mean cluster parallaxes could be quite
complicated.
For example, another nearby open cluster NGC~2451A, which is located only
$18\degr$ away from IC~2391, does not show any discrepancy between
the Hipparcos parallax and the distance modulus derived from the main sequence
fitting \citep{pla01}, while the number of measured stars and their
properties in both clusters are nearly identical. Both clusters are
located near the ecliptic latitude of $\beta\sim-60\degr$ and
had been observed by Hipparcos relatively frequently and close in
time. Comparative studies of these two open clusters may reveal clues
to the cause of a $\sim$0.5 mas offset in the Hipparcos mean
parallax for IC~2391.

Among the most important tasks to advance the knowledge about
IC~2391 are 1) extending reliable kinematic membership to fainter
magnitudes and 2) accumulating high-resolution spectroscopy for all
possible cluster members, especially for known or suspected 
spectroscopic binaries. The cluster population should be cleanly
separated into single and binaries stars prior to extensive
studies of individual stars and establishing a reliable distance to
IC~2391 by other means.

\begin{acknowledgements} 

We thank Hugo Levato for organizing the observing run at the CASLEO
2.2~m telescope. We appreciate helpful comments by Heiner Schwan and
Peter De Cat about the properties of $o$~Velorum. 
We also thank Jeremy King and Simon Schuler for stimulating
discussions on elemental abundances. 
We thank the referee, John Stauffer,
for detailed and thoughtful comments that greatly helped to improve the
interpretation of the results. We also thank Deokkeun An, Don Terndrup,
and Marc Pinsonneault for their help with the new YREC isochrones generated
at Ohio State University.  
This research made use of the SIMBAD
database operated at the CDS, Strasbourg, France. This publication made use 
of data products from the Two Micron All Sky Survey, which is 
a joint project of the University of Massachusetts and the Infrared 
Processing and Analysis Center/California Institute of Technology, 
funded by the National Aeronautics and Space Administration and 
the National Science Foundation. 
I. Platais gratefully acknowledges
support from the National Science Foundation through grant AST
04-06689 to Johns Hopkins University.
J. Fulbright acknowledges support through grants from the W.M. Keck
Foundation and the Gordon and Betty Moore Foundation, to establish
a program of data-intensive science at to the Johns Hopkins University.
M. Altmann and R. Mendez acknowledge the support by the Chilean Centro
de Astrof\'{i}sica FONDAP (15010003). The travel by J. Sperauskas to
El Leoncito, Argentina was in part supported by NSF supplemental funding
AST 01-39797. We heartily thank Lois J. Evans for meticulous  editing of
this paper.

  \end{acknowledgements}

\bibliographystyle{aa}
\bibliography{5756bib}

\begin{thebibliography}{96}
\expandafter\ifx\csname natexlab\endcsname\relax\def\natexlab#1{#1}\fi

\bibitem[{{Allende Prieto} {et~al.}(2004){Allende Prieto}, {Barklem},
  {Lambert}, \& {Cunha}}]{all04}
{Allende Prieto}, C., {Barklem}, P.~S., {Lambert}, D.~L., \& {Cunha}, K. 2004,
  \aap, 420, 183

\bibitem[{{Aller} {et~al.}(1982){Aller}, {Appenzeller}, {Baschek}, {Duerbeck},
  {Herczeg}, {Lamla}, {Meyer-Hofmeister}, {Schmidt-Kaler}, {Scholz},
  {Seggewiss}, {Seitter}, \& {Weidemann}}]{lan82}
{Aller}, L.~H., {Appenzeller}, I., {Baschek}, B., {et~al.} 1982,
  {Landolt-B{\"o}rnstein: Numerical Data and Functional Relationships in
  Science and Technology - New Series '' Group 6 Astronomy and Astrophysics ''
  Volume 2 '' } (Landolt-Bornstein: Numerical Data and Functional Relationships
  in Science and Technology)

\bibitem[{{An} {et~al.}(2006){An}, {Terndrup}, {Pinsonneault}, {Paulson},
  {Hanson}, , \& {Stauffer}}]{an06}
{An}, D., {Terndrup}, D.~M., {Pinsonneault}, M.~H., {et~al.} 2006, {ApJ,
  submitted (astro-ph/0607549)}

\bibitem[{{Baranne} {et~al.}(1979){Baranne}, {Mayor}, \& {Poncet}}]{bar79}
{Baranne}, A., {Mayor}, M., \& {Poncet}, J.~L. 1979, Vistas in Astronomy, 23,
  279

\bibitem[{{Baranne} {et~al.}(1996){Baranne}, {Queloz}, {Mayor}, {Adrianzyk},
  {Knispel}, {Kohler}, {Lacroix}, {Meunier}, {Rimbaud}, \& {Vin}}]{bar96}
{Baranne}, A., {Queloz}, D., {Mayor}, M., {et~al.} 1996, \aaps, 119, 373

\bibitem[{{Barnes} \& {Sofia}(1996)}]{bars96}
{Barnes}, S. \& {Sofia}, S. 1996, \apj, 462, 746

\bibitem[{{Barnes}(2003)}]{bar03}
{Barnes}, S.~A. 2003, \apj, 586, 464

\bibitem[{{Barrado y Navascu{\' e}s} {et~al.}(1999){Barrado y Navascu{\' e}s},
  {Stauffer}, \& {Patten}}]{bar99}
{Barrado y Navascu{\' e}s}, D., {Stauffer}, J.~R., \& {Patten}, B.~M. 1999,
  \apjl, 522, L53

\bibitem[{{Barrado y Navascu{\'e}s} {et~al.}(2004){Barrado y Navascu{\'e}s},
  {Stauffer}, \& {Jayawardhana}}]{bar04}
{Barrado y Navascu{\'e}s}, D., {Stauffer}, J.~R., \& {Jayawardhana}, R. 2004,
  \apj, 614, 386

\bibitem[{{Benz} \& {Mayor}(1981)}]{ben81}
{Benz}, W. \& {Mayor}, M. 1981, \aap, 93, 235

\bibitem[{{Benz} \& {Mayor}(1984)}]{ben84}
{Benz}, W. \& {Mayor}, M. 1984, \aap, 138, 183

\bibitem[{{Blair} \& {Gilmore}(1982)}]{bla82}
{Blair}, M. \& {Gilmore}, G. 1982, \pasp, 94, 742

\bibitem[{{Bonatto} {et~al.}(2004){Bonatto}, {Bica}, \& {Girardi}}]{bon04}
{Bonatto}, C., {Bica}, E., \& {Girardi}, L. 2004, \aap, 415, 571

\bibitem[{{Buscombe}(1965)}]{bus65}
{Buscombe}, W. 1965, \mnras, 129, 411

\bibitem[{{Cutri} {et~al.}(2003){Cutri}, {Skrutskie}, {van Dyk}, {Beichman},
  {Carpenter}, {Chester}, {Cambresy}, {Evans}, {Fowler}, {Gizis}, {Howard},
  {Huchra}, {Jarrett}, {Kopan}, {Kirkpatrick}, {Light}, {Marsh}, {McCallon},
  {Schneider}, {Stiening}, {Sykes}, {Weinberg}, {Wheaton}, {Wheelock}, \&
  {Zacharias}}]{cut03}
{Cutri}, R.~M., {Skrutskie}, M.~F., {van Dyk}, S., {et~al.} 2003, {2MASS All
  Sky Catalog of point sources.} (The IRSA 2MASS All-Sky Point Source Catalog,
  NASA/IPAC Infrared Science
  Archive.~http://irsa.ipac.caltech.edu/applications/Gator/)

\bibitem[{{Dodd}(2004)}]{dod04}
{Dodd}, R.~J. 2004, \mnras, 355, 959

\bibitem[{{Edwards} {et~al.}(1993){Edwards}, {Strom}, {Hartigan}, {Strom},
  {Hillenbrand}, {Herbst}, {Attridge}, {Merrill}, {Probst}, \&
  {Gatley}}]{edw93}
{Edwards}, S., {Strom}, S.~E., {Hartigan}, P., {et~al.} 1993, \aj, 106, 372

\bibitem[{{Eggen}(1991)}]{egg91}
{Eggen}, O.~J. 1991, \aj, 102, 2028

\bibitem[{{Feinstein}(1961)}]{fei61}
{Feinstein}, A. 1961, \pasp, 73, 410

\bibitem[{{Forbes} {et~al.}(1998){Forbes}, {Dodd}, \& {Sullivan}}]{for98}
{Forbes}, M.~C., {Dodd}, R.~J., \& {Sullivan}, D.~J. 1998, Baltic Astronomy,
  10, 375

\bibitem[{{Fox Machado} {et~al.}(2006){Fox Machado}, {P{\'e}rez Hern{\'a}ndez},
  {Su{\'a}rez}, {Michel}, \& {Lebreton}}]{fox06}
{Fox Machado}, L., {P{\'e}rez Hern{\'a}ndez}, F., {Su{\'a}rez}, J.~C.,
  {Michel}, E., \& {Lebreton}, Y. 2006, \aap, 446, 611

\bibitem[{{Fulbright} {et~al.}(2006){Fulbright}, {McWilliam}, \&
  {Rich}}]{ful06}
{Fulbright}, J.~P., {McWilliam}, A., \& {Rich}, R.~M. 2006, \apj, 636, 821

\bibitem[{{Gatewood} {et~al.}(2000){Gatewood}, {de Jonge}, \& {Han}}]{gat00}
{Gatewood}, G., {de Jonge}, J.~K., \& {Han}, I. 2000, \apj, 533, 938

\bibitem[{{Girard} {et~al.}(1998){Girard}, {Platais}, {Kozhurina-Platais}, {van
  Altena}, \& {L{\'o}pez}}]{gir98}
{Girard}, T.~M., {Platais}, I., {Kozhurina-Platais}, V., {van Altena}, W.~F.,
  \& {L{\'o}pez}, C.~E. 1998, \aj, 115, 855

\bibitem[{{Girardi} {et~al.}(2002){Girardi}, {Bertelli}, {Bressan}, {Chiosi},
  {Groenewegen}, {Marigo}, {Salasnich}, \& {Weiss}}]{gir02}
{Girardi}, L., {Bertelli}, G., {Bressan}, A., {et~al.} 2002, \aap, 391, 195

\bibitem[{{Gontcharov} {et~al.}(2001){Gontcharov}, {Andronova}, {Titov}, \&
  {Kornilov}}]{gon01}
{Gontcharov}, G.~A., {Andronova}, A.~A., {Titov}, O.~A., \& {Kornilov}, E.~V.
  2001, \aap, 365, 222

\bibitem[{{Gray}(1992)}]{Gray92}
{Gray}, D.~F. 1992, {The observation and analysis of stellar photospheres}
  (Cambridge Astrophysics Series, Cambridge: Cambridge University Press, 1992,
  2nd ed., ISBN 0521403200.)

\bibitem[{{Haisch} {et~al.}(2001){Haisch}, {Lada}, \& {Lada}}]{hai01}
{Haisch}, K.~E., {Lada}, E.~A., \& {Lada}, C.~J. 2001, \apjl, 553, L153

\bibitem[{{Herbst} \& {Mundt}(2005)}]{her05}
{Herbst}, W. \& {Mundt}, R. 2005, \apj, 633, 967

\bibitem[{{Hogg}(1960)}]{hog60}
{Hogg}, A.~R. 1960, \pasp, 72, 85

\bibitem[{{Jeffries} \& {Oliveira}(2005)}]{jef05}
{Jeffries}, R.~D. \& {Oliveira}, J.~M. 2005, \mnras, 358, 13

\bibitem[{{Kaufer} {et~al.}(1999){Kaufer}, {Stahl}, {Tubbesing}, {Norregaard},
  {Avila}, {Francois}, {Pasquini}, \& {Pizzella}}]{kaufer99}
{Kaufer}, A., {Stahl}, O., {Tubbesing}, S., {et~al.} 1999, The Messenger, 95, 8

\bibitem[{{King}(1979)}]{kin79}
{King}, D.~S. 1979, Journal and Proceedings of the Royal Society of New South
  Wales, 112, 105

\bibitem[{{Kozhurina-Platais} {et~al.}(1995){Kozhurina-Platais}, {Girard},
  {Platais}, {van Altena}, {Ianna}, \& {Cannon}}]{koz95}
{Kozhurina-Platais}, V., {Girard}, T.~M., {Platais}, I., {et~al.} 1995, \aj,
  109, 672

\bibitem[{{Kurucz} {et~al.}(1984){Kurucz}, {Furenlid}, \& {Brault}}]{kur84}
{Kurucz}, R.~L., {Furenlid}, I., \& {Brault}, J.~T.~L. 1984, {Solar flux atlas
  from 296 to 1300 nm} (National Solar Observatory Atlas, Sunspot, New Mexico:
  National Solar Observatory, 1984)

\bibitem[{{Landolt}(1992)}]{lan92}
{Landolt}, A.~U. 1992, \aj, 104, 340

\bibitem[{{Lee} \& {van Altena}(1983)}]{lee83}
{Lee}, J.-F. \& {van Altena}, W. 1983, \aj, 88, 1683

\bibitem[{{Lejeune} {et~al.}(1998){Lejeune}, {Cuisinier}, \& {Buser}}]{lej98}
{Lejeune}, T., {Cuisinier}, F., \& {Buser}, R. 1998, \aaps, 130, 65

\bibitem[{{Levato} {et~al.}(1988){Levato}, {Garcia}, {Lousto}, \&
  {Morrell}}]{lev88}
{Levato}, H., {Garcia}, B., {Lousto}, C., \& {Morrell}, N. 1988, \apss, 146,
  361

\bibitem[{{Lyng{\aa}}(1960)}]{lyn60}
{Lyng{\aa}}, G. 1960, Arkiv for Astronomi, 2, 379

\bibitem[{{Makarov}(2002)}]{mak02}
{Makarov}, V.~V. 2002, \aj, 124, 3299

\bibitem[{{Marino} {et~al.}(2005){Marino}, {Micela}, {Peres}, {Pillitteri}, \&
  {Sciortino}}]{mar05}
{Marino}, A., {Micela}, G., {Peres}, G., {Pillitteri}, I., \& {Sciortino}, S.
  2005, \aap, 430, 287

\bibitem[{{Marino} {et~al.}(2003){Marino}, {Micela}, {Peres}, \&
  {Sciortino}}]{mar03}
{Marino}, A., {Micela}, G., {Peres}, G., \& {Sciortino}, S. 2003, \aap, 407,
  L63

\bibitem[{{Marsden} {et~al.}(2004){Marsden}, {Waite}, {Carter}, \&
  {Donati}}]{mars04}
{Marsden}, S.~C., {Waite}, I.~A., {Carter}, B.~D., \& {Donati}, J.-F. 2004,
  Astronomische Nachrichten, 325, 246

\bibitem[{{Martin}(1997)}]{martin97}
{Martin}, E.~L. 1997, \aap, 321, 492

\bibitem[{{Mathieu}(2000)}]{mat00}
{Mathieu}, R.~D. 2000, in ASP Conf. Ser. 198: Stellar Clusters and
  Associations: Convection, Rotation, and Dynamos, ed. R.~{Pallavicini},
  G.~{Micela}, \& S.~{Sciortino}, 517

\bibitem[{{Mayor}(1985)}]{may85}
{Mayor}, M. 1985, in IAU Colloq. 88: Stellar Radial Velocities, ed. A.~G.~D.
  {Philip} \& D.~W. {Latham}, 35

\bibitem[{{Melo}(2003)}]{melo03}
{Melo}, C.~H.~F. 2003, \aap, 410, 269

\bibitem[{{Melo} {et~al.}(2001){Melo}, {Pasquini}, \& {De Medeiros}}]{melo01}
{Melo}, C.~H.~F., {Pasquini}, L., \& {De Medeiros}, J.~R. 2001, \aap, 375, 851

\bibitem[{{Mermilliod}(1981)}]{mer81}
{Mermilliod}, J.~C. 1981, \aap, 97, 235

\bibitem[{{Mermilliod} \& {Mayor}(1989)}]{mer89}
{Mermilliod}, J.-C. \& {Mayor}, M. 1989, \aap, 219, 125

\bibitem[{{Munari} {et~al.}(2004){Munari}, {Dallaporta}, {Siviero}, {Soubiran},
  {Fiorucci}, \& {Girard}}]{mun04}
{Munari}, U., {Dallaporta}, S., {Siviero}, A., {et~al.} 2004, \aap, 418, L31

\bibitem[{{Nordstroem} {et~al.}(1996){Nordstroem}, {Andersen}, \&
  {Andersen}}]{nor96}
{Nordstroem}, B., {Andersen}, J., \& {Andersen}, M.~I. 1996, \aaps, 118, 407

\bibitem[{{Pan} {et~al.}(2004){Pan}, {Shao}, \& {Kulkarni}}]{pan04}
{Pan}, X., {Shao}, M., \& {Kulkarni}, S.~R. 2004, \nat, 427, 326

\bibitem[{{Patten} \& {Simon}(1993)}]{pat93}
{Patten}, B.~M. \& {Simon}, T. 1993, \apjl, 415, L123

\bibitem[{{Patten} \& {Simon}(1996)}]{pat96}
{Patten}, B.~M. \& {Simon}, T. 1996, \apjs, 106, 489

\bibitem[{{Paulson} {et~al.}(2004){Paulson}, {Cochran}, \& {Hatzes}}]{pau04}
{Paulson}, D.~B., {Cochran}, W.~D., \& {Hatzes}, A.~P. 2004, \aj, 127, 3579

\bibitem[{{Paulson} {et~al.}(2003){Paulson}, {Sneden}, \& {Cochran}}]{pau03}
{Paulson}, D.~B., {Sneden}, C., \& {Cochran}, W.~D. 2003, \aj, 125, 3185

\bibitem[{{Perry} \& {Bond}(1969)}]{per69a}
{Perry}, C.~L. \& {Bond}, H.~E. 1969, \pasp, 81, 629

\bibitem[{{Perry} \& {Hill}(1969)}]{per69b}
{Perry}, C.~L. \& {Hill}, G. 1969, \aj, 74, 899

\bibitem[{{Pinsonneault} {et~al.}(1998){Pinsonneault}, {Stauffer}, {Soderblom},
  {King}, \& {Hanson}}]{pin98}
{Pinsonneault}, M.~H., {Stauffer}, J., {Soderblom}, D.~R., {King}, J.~R., \&
  {Hanson}, R.~B. 1998, \apj, 504, 170

\bibitem[{{Pinsonneault} {et~al.}(2003){Pinsonneault}, {Terndrup}, {Hanson}, \&
  {Stauffer}}]{pin03}
{Pinsonneault}, M.~H., {Terndrup}, D.~M., {Hanson}, R.~B., \& {Stauffer}, J.~R.
  2003, \apj, 598, 588

\bibitem[{{Pinsonneault} {et~al.}(2004){Pinsonneault}, {Terndrup}, {Hanson}, \&
  {Stauffer}}]{pin04}
{Pinsonneault}, M.~H., {Terndrup}, D.~M., {Hanson}, R.~B., \& {Stauffer}, J.~R.
  2004, \apj, 600, 946

\bibitem[{{Platais} {et~al.}(1998){Platais}, {Girard}, {Kozhurina-Platais},
  {van Altena}, {L{\'o}pez}, {M{\'e}ndez}, {Ma}, {Yang}, {MacGillivray}, \&
  {Yentis}}]{pla98}
{Platais}, I., {Girard}, T.~M., {Kozhurina-Platais}, V., {et~al.} 1998, \aj,
  116, 2556

\bibitem[{{Platais} {et~al.}(2001){Platais}, {Kozhurina-Platais}, {Barnes},
  {Girard}, {Demarque}, {van Altena}, {Deliyannis}, \& {Horch}}]{pla01}
{Platais}, I., {Kozhurina-Platais}, V., {Barnes}, S., {et~al.} 2001, \aj, 122,
  1486

\bibitem[{{Prosser}(1992)}]{pro92}
{Prosser}, C.~F. 1992, \aj, 103, 488

\bibitem[{Queloz(1995)}]{queloz95}
Queloz, D. 1995, PhD thesis, Universit\'{e} de Gen\`{e}ve

\bibitem[{{Ram{\'{\i}}rez} \& {Mel{\'e}ndez}(2005)}]{ram05}
{Ram{\'{\i}}rez}, I. \& {Mel{\'e}ndez}, J. 2005, \apj, 626, 465

\bibitem[{{Randich} {et~al.}(2001){Randich}, {Pallavicini}, {Meola},
  {Stauffer}, \& {Balachandran}}]{ran01}
{Randich}, S., {Pallavicini}, R., {Meola}, G., {Stauffer}, J.~R., \&
  {Balachandran}, S.~C. 2001, \aap, 372, 862

\bibitem[{{Robichon} {et~al.}(1999){Robichon}, {Arenou}, {Mermilliod}, \&
  {Turon}}]{rob99}
{Robichon}, N., {Arenou}, F., {Mermilliod}, J.-C., \& {Turon}, C. 1999, \aap,
  345, 471

\bibitem[{{Schuler} {et~al.}(2006){Schuler}, {King}, {Terndrup},
  {Pinsonneault}, {Murray}, \& {Hobbs}}]{sch06}
{Schuler}, S.~C., {King}, J.~R., {Terndrup}, D.~M., {et~al.} 2006, \apj, 636,
  432

\bibitem[{{Setiawan} {et~al.}(2000){Setiawan}, {Pasquini}, {da Silva},
  {Hatzes}, {von der Luhe}, {Kaufer}, {Girardi}, {de La}, \& {de
  Medeiros}}]{setiawan00}
{Setiawan}, J., {Pasquini}, L., {da Silva}, L., {et~al.} 2000, The Messenger,
  102, 13

\bibitem[{{Simon} \& {Patten}(1998)}]{sim98}
{Simon}, T. \& {Patten}, B.~M. 1998, \pasp, 110, 283

\bibitem[{{Skrutskie} {et~al.}(2006){Skrutskie}, {Cutri}, {Stiening},
  {Weinberg}, {Schneider}, {Carpenter}, {Beichman}, {Capps}, {Chester},
  {Elias}, {Huchra}, {Liebert}, {Lonsdale}, {Monet}, {Price}, {Seitzer},
  {Jarrett}, {Kirkpatrick}, {Gizis}, {Howard}, {Evans}, {Fowler}, {Fullmer},
  {Hurt}, {Light}, {Kopan}, {Marsh}, {McCallon}, {Tam}, {Van Dyk}, \&
  {Wheelock}}]{skr06}
{Skrutskie}, M.~F., {Cutri}, R.~M., {Stiening}, R., {et~al.} 2006, \aj, 131,
  1163

\bibitem[{{Sneden}(1973)}]{moog}
{Sneden}, C.~A. 1973, Ph.D.~Thesis

\bibitem[{{Soderblom} {et~al.}(1993){Soderblom}, {Jones}, {Balachandran},
  {Stauffer}, {Duncan}, {Fedele}, \& {Hudon}}]{sod93}
{Soderblom}, D.~R., {Jones}, B.~F., {Balachandran}, S., {et~al.} 1993, \aj,
  106, 1059

\bibitem[{{Soderblom} {et~al.}(2005){Soderblom}, {Nelan}, {Benedict},
  {McArthur}, {Ramirez}, {Spiesman}, \& {Jones}}]{sod05}
{Soderblom}, D.~R., {Nelan}, E., {Benedict}, G.~F., {et~al.} 2005, \aj, 129,
  1616

\bibitem[{{Stauffer} {et~al.}(1989){Stauffer}, {Hartmann}, {Jones}, \&
  {McNamara}}]{sta89}
{Stauffer}, J., {Hartmann}, L.~W., {Jones}, B.~F., \& {McNamara}, B.~R. 1989,
  \apj, 342, 285

\bibitem[{{Stauffer} {et~al.}(1999){Stauffer}, {Barrado y Navascu{\'e}s},
  {Bouvier}, {Morrison}, {Harding}, {Luhman}, {Stanke}, {McCaughrean},
  {Terndrup}, {Allen}, \& {Assouad}}]{sta99}
{Stauffer}, J.~R., {Barrado y Navascu{\'e}s}, D., {Bouvier}, J., {et~al.} 1999,
  \apj, 527, 219

\bibitem[{{Stauffer} {et~al.}(1997){Stauffer}, {Hartmann}, {Prosser},
  {Randich}, {Balachandran}, {Patten}, {Simon}, \& {Giampapa}}]{sta97}
{Stauffer}, J.~R., {Hartmann}, L.~W., {Prosser}, C.~F., {et~al.} 1997, \apj,
  479, 776

\bibitem[{{Stauffer} {et~al.}(2003){Stauffer}, {Jones}, {Backman}, {Hartmann},
  {Barrado y Navascu{\'e}s}, {Pinsonneault}, {Terndrup}, \& {Muench}}]{sta03}
{Stauffer}, J.~R., {Jones}, B.~F., {Backman}, D., {et~al.} 2003, \aj, 126, 833

\bibitem[{{Stauffer} {et~al.}(1998){Stauffer}, {Schultz}, \&
  {Kirkpatrick}}]{sta98}
{Stauffer}, J.~R., {Schultz}, G., \& {Kirkpatrick}, J.~D. 1998, \apjl, 499,
  L199

\bibitem[{{St{\"u}tz} {et~al.}(2006){St{\"u}tz}, {Bagnulo}, {Jehin}, {Ledoux},
  {Cabanac}, {Melo}, \& {Smoker}}]{stu06}
{St{\"u}tz}, C., {Bagnulo}, S., {Jehin}, E., {et~al.} 2006, \aap, 451, 285

\bibitem[{{Udry} {et~al.}(1999){Udry}, {Mayor}, \& {Queloz}}]{udr99}
{Udry}, S., {Mayor}, M., \& {Queloz}, D. 1999, in ASP Conf. Ser. 185: IAU
  Colloq. 170: Precise Stellar Radial Velocities, ed. J.~B. {Hearnshaw} \&
  C.~D. {Scarfe}, 367

\bibitem[{{Upgren} {et~al.}(2002){Upgren}, {Sperauskas}, \& {Boyle}}]{upg02}
{Upgren}, A.~R., {Sperauskas}, J., \& {Boyle}, R.~P. 2002, Baltic Astronomy,
  11, 91

\bibitem[{{van Hoof}(1972)}]{vhoo72}
{van Hoof}, A. 1972, \aap, 18, 51

\bibitem[{{van Leeuwen}(1999)}]{vlee99}
{van Leeuwen}, F. 1999, \aap, 341, L71

\bibitem[{{van Leeuwen}(2005)}]{vlee05a}
{van Leeuwen}, F. 2005, in IAU Colloq. 196: Transits of Venus: New Views of the
  Solar System and Galaxy, ed. D.~W. {Kurtz}, 347--360

\bibitem[{{van Leeuwen} {et~al.}(1987){van Leeuwen}, {Alphenaar}, \&
  {Meys}}]{vlee87}
{van Leeuwen}, F., {Alphenaar}, P., \& {Meys}, J.~J.~M. 1987, \aaps, 67, 483

\bibitem[{{van Leeuwen} \& {Fantino}(2005)}]{vlee05b}
{van Leeuwen}, F. \& {Fantino}, E. 2005, \aap, 439, 791

\bibitem[{{Wielen} {et~al.}(1999){Wielen}, {Dettbarn}, {Jahrei{\ss}},
  {Lenhardt}, \& {Schwan}}]{wie99}
{Wielen}, R., {Dettbarn}, C., {Jahrei{\ss}}, H., {Lenhardt}, H., \& {Schwan},
  H. 1999, \aap, 346, 675

\bibitem[{{Wilson}(1941)}]{wil41}
{Wilson}, O.~C. 1941, \apj, 93, 29

\bibitem[{{Yentis} {et~al.}(1992){Yentis}, {Cruddace}, {Gursky}, {Stuart},
  {Wallin}, {MacGillivray}, \& {Collins}}]{yen92}
{Yentis}, D.~J., {Cruddace}, R.~G., {Gursky}, H., {et~al.} 1992, in ASSL Vol.
  174: Digitised Optical Sky Surveys, ed. H.~T. {MacGillivray} \& E.~B.
  {Thomson}, 67

\bibitem[{{Yong} {et~al.}(2004){Yong}, {Lambert}, {Allende Prieto}, \&
  {Paulson}}]{yon04}
{Yong}, D., {Lambert}, D.~L., {Allende Prieto}, C., \& {Paulson}, D.~B. 2004,
  \apj, 603, 697

\bibitem[{{Zacharias} {et~al.}(2004){Zacharias}, {Urban}, {Zacharias},
  {Wycoff}, {Hall}, {Monet}, \& {Rafferty}}]{zac04}
{Zacharias}, N., {Urban}, S.~E., {Zacharias}, M.~I., {et~al.} 2004, \aj, 127,
  3043

\bibitem[{{Zwahlen} {et~al.}(2004){Zwahlen}, {North}, {Debernardi}, {Eyer},
  {Galland}, {Groenewegen}, \& {Hummel}}]{zwa04}
{Zwahlen}, N., {North}, P., {Debernardi}, Y., {et~al.} 2004, \aap, 425, L45

\end{thebibliography}


\newpage
\begin{table*}[htb]
\setcounter{table}{1}
\caption[]{\label{tab:phot}Comparison of $BV$ photometry `CCD$-$Source'}
\begin{tabular}{lccc}\hline\hline
Source & n & $\Delta V$ & $\Delta (B-V)$\\ \hline

Hogg (1960) & 24 & $-0.025\pm 0.07$ & $+0.016\pm 0.03$ \\
Lyng\aa~(1960) & 24 & $+0.040\pm 0.06$ & $-0.011\pm 0.03$ \\
Perry \& Hill (1969) & 21 & $+0.013\pm 0.05$ & $+0.012\pm 0.04$ \\
Forbes et al. (1998) & 19 & $+0.038\pm 0.06$ & \nodata\\
Eggen (1972) & 11 & $+0.050\pm 0.06$ & $-0.010\pm 0.02$\\
Stauffer et al. (1989) & 5 & $+0.058\pm 0.05$ & $+0.046\pm 0.04$\\
Rollestone \& Byrne (1997) & 4 & $+0.027\pm 0.02$ & $+0.14\pm 0.03$\\
\hline
\end{tabular}
\end{table*}


\newpage
\setcounter{table}{3}
\begin{table*}[h]
\begin{center}
\caption[]{\label{tab:meanrv}Mean radial velocities and vsini from Coravel observations}
\begin{tabular}{rrccrrrrrrl}
\hline
No &  $<$RV$>$ & $\sigma$  &  $\epsilon$ &  E/I &  $n$ &  $\Delta$T &  $v\sin i$ & $\sigma_{v\sin i}$  & P($\chi^2$) &  Remarks\\
\hline\hline
 389  &  9.27 & 2.75 & 1.38 & 4.69 & 4 & 4412 & 14.4 & 1.8 & 0.000 & SB\\
 \nodata  & 5.00 & 8.35 & 2.52 & 10.49 & 11 & 3250 & 8.0 & 3.4 & 0.000 & SB2 (A)\\
 \nodata  &  17.00 & 10.83&  3.27 & 7.39 & 11 & 3250 & 3.0 & 3.7 & 0.000 & SB2 (B)\\
 736  &  -25.11 & 3.17 & 3.17 & 1.00 & 1 &    0 & 22.1 & 11.2 & 9.999 & SB2 (A)\\
 \nodata  &  21.53 & 3.23 & 3.23 & 1.00 & 1 &    0 & 41.2 & 15.0 & 9.999 & SB2 (B)\\
1083  &  12.17 & 0.57 & 1.04 & 0.38 & 2 &  323 &  67.9 & 6.8 & 0.702 & \\
1142  &  13.99 & 0.42 & 0.42 & 1.00 & 1 &    0 & 7.9 & 1.7 & 9.999 & \\
1759  &  13.51 & 0.51 & 0.51 & 1.00 & 1 &    0 & 3.8 & 3.5 & 9.999 & \\
2012  &  13.90 & 0.63 & 0.63 & 1.00 & 1 &    0 & 17.4 & 1.6 & 9.999 & \\
2888  &  14.81 & 2.68 & 1.55 & 1.00 & 3 &  740 & 66.5 & 6.7 & 0.382 & \\
3464  &  13.93 & 2.77 & 1.05 & 2.78 & 7 & 4412 & 20.8 & 2.1 & 0.000 & SB\\
3497  &  31.69 & 0.65 & 0.65 & 1.00 & 1 &    0 & 11.8 & 2.5 & 9.999 & \\
3567  &  13.33 & 0.37 & 0.37 & 1.00 & 1 &    0 & 3.9 & 2.5 & 9.999 & \\
3623  &  24.05 & 0.37 & 0.37 & 1.00 & 1 &    0 & 4.2 & 2.3 & 9.999 & \\
3649  &  15.85 & 0.88 & 0.88 & 1.00 & 1 &    0 & 19.2 & 3.1 & 9.999 & \\
3664  &  6.70 & 2.28 & 1.14 & 4.42 & 4 & 2133 & 3.8 & 1.9 & 0.000 & SB \\
3722  &  15.25 & 4.74 & 1.43 & 9.33 & 11 & 4406 & 10.1 & 0.6 & 0.000 & SB \\
4336  &  15.63 & 0.41 & 0.41 & 1.00 & 1 &    0 & 8.0 & 1.9 & 9.999 & \\
4362  &  15.20 & 0.51 & 0.17 & 1.04 & 9 & 4412 & 9.0 & 0.7 & 0.401 & \\
4413  &  3.32 & 16.36& 3.57  & 22.53 & 21 & 4412 & 8.6 & 0.7 & 0.000 & SB2 (A)\\
\nodata  &  25.92 & 16.74&  3.65 & 21.24 & 21 & 4412 & 8.4 & 0.7 & 0.000 & SB2 (B)\\
4467  &  15.22 & 0.19 & 0.32 & 0.33 & 3 & 2225 & 12.6 & 1.0 & 0.894 & \\
4549  &  16.59 & 17.81&  6.73 & 8.31 & 7 & 4413 & 48.8 & 6.2 & 0.000 & SB \\
4636  &  13.63 & 0.52 & 0.52 & 1.00 & 1 &    0 & 4.9 & 2.9 & 9.999 & \\
4809  &  13.70 & 0.17 & 0.45 & 0.27 & 2 &  419 & 17.4 & 1.1 & 0.790 & \\
4902  &  14.71 & 0.59 & 0.42 & 1.01 & 2 &  418 & 8.3 & 1.8 & 0.313 & \\
5382  &  5.69 & 3.34 & 1.06 & 7.69 & 10 & 4411 & 8.7 & 0.8 & 0.000 & SB\\
5768  &  15.21 & 3.90 & 2.76 & 3.76 & 2 &    5 & 27.3 & 2.7 & 0.000 & SB\\
5829  &  10.03 & 1.65 & 1.65 & 1.00 & 1 &    0 & 67.6 & 9.3 & 9.999 & \\
5859  &  17.46 & 2.49 & 1.76 & 5.50 & 2 &  414 & 8.5 & 2.3 & 0.000 & SB\\
5884  &  14.49 & 0.56 & 0.40 & 1.01 & 2 &  419 & 13.9 & 1.1 & 0.313 & \\
6478  &  21.98 & 1.41 & 1.00 & 2.09 & 2 &  322 & 20.4 & 2.0 & 0.038 & \\
7794  &   9.38 & 0.57 & 0.37 & 0.58 & 7 & 4412 & 21.3 & 2.1 & 0.921 & \\
8415  &  15.73 & 0.66 & 0.66 & 1.00 & 1 &    0 & 20.7 & 2.3 & 9.999 & \\
\hline
\end{tabular}
\end{center}
\end{table*}



\newpage
\setcounter{table}{6}
\begin{table*}[htb]
\caption[]{\label{tab:abund}Derived stellar atmosphere parameters and Fe abundance}
\begin{tabular}{rcccccccccc} \hline\hline
Star & T$_{\rm eff}$ & $\log g$ & [m/H]$^{\mathrm{a}}$ & v$_{\rm t}$ & log $\epsilon$(Fe~I)& $\sigma$ & N & log $\epsilon$(Fe~II)& $\sigma$ &N\\
 & ($^\circ$K) & & & (km s$^{-1}$) &  &  &  &  &  & \\ \hline
 665 & 5451 & 4.53 & $+0.05$ & 1.25 & 7.50 & 0.06 & 64 & 7.60 & 0.07 & 5 \\
1560 & 5830 & 4.48 & $+0.13$ & 1.38 & 7.58 & 0.07 & 76 & 7.87 & 0.03 & 5 \\
3359 & 5280 & 4.55 & $-0.01$ & 1.17 & 7.44 & 0.06 & 73 & 7.64 & 0.06 & 4 \\
4362 & 5616 & 4.51 & $+0.05$ & 1.30 & 7.50 & 0.07 & 67 & 7.59 & 0.04 & 5 \\
\hline
\end{tabular}
\begin{list}{}{}
\item[$^{\mathrm{a}}$] Input model atmosphere abundance on a scale where 
the solar $\log$ $\epsilon$(Fe)$= 7.45$.
\end{list}
\end{table*}

\newpage
\setcounter{table}{7}
\begin{table*}[htb]
\caption[]{\label{tab:meanli}Mean data from FEROS spectroscopy}
\setlength{\tabcolsep}{1mm}
\begin{tabular}{rcccccc|rcccccc}\hline \hline
No & $<$RV$>$ & $v\sin i$ & $T_{\rm eff}$ &  EW(Li) & log N(Li) & EW(H$_\alpha$) &
No & $<$RV$>$ & $v\sin i$ & $T_{\rm eff}$ &  EW(Li) & log N(Li) & EW(H$_\alpha$)\\
   &  (km s$^{-1}$) & (km s$^{-1}$) & ($^{\circ}$K) & (m\AA) & & (\AA) &
   &  (km s$^{-1}$) & (km s$^{-1}$) & ($^{\circ}$K) & (m\AA) & & (\AA) \\ \hline 
 351	&	  15.40	&	       90.0 &  6024 &	 90 & 2.82 &  3.69 & 4549    &         20.21 &  	    39.0 &  6192 &   150 & 3.36 &    2.9 \\
 665	&	  14.44	&	        8.3 &  5451 &	189 & 2.88 &  2.11 & 4809    &         13.80 &  	    17.7 &  5546 &   175 & 2.90 &    3.1 \\
 686	&	  15.18	&	       12.7 &  4538 &	215 & 1.87 & -0.13 & 5137    &         25.38 &  	     3.4 &  5148 &     0 & \nodata &	1.5 \\
 736	&	  13.54	&	       65.0 &  6579 &	 83 & 3.22 &  5.10 & 5314    &         13.36 &  	    63.0 &  6467 &    -99 & \nodata &	 5.62\\
 756	&	  15.86	&	       16.5 &  5357 &	217 & 2.96 &  2.7  & 5376    &         15.62 &  	     9.8 &  3936 &     0 & \nodata &   -1.17\\
 794	&	  16.41	&	        2.6 &  4874 &	  0 & \nodata&  1.4& 5540    &         54.47 &  	     6.5 &  5652 &    48 & 2.13 &    3.10\\
 819	&	  17.84	&	       30.0 &  5764 &	179 & 3.15 &  3.7  & 5811    &         14.30 &  	    59.0 &  6823 &   107 & 3.58 &    4.84\\
1174	&	  13.54	&	       60.0 &  6584 &	 79 & 3.20 &  3.48 & 5859    &         16.86 &  	     8.8 &  4749 &   271 & 2.54 &   -0.07\\
1373	&	  14.62	&	        7.4 &  4746 &	156 & 1.83 &  1.0  & 6229    &	       11.73 &               3.7 &  5200 &    61 & 1.78 &     2.1\\
1560	&	  15.98	&	        5.7 &  5830 &	107 & 2.77 &  2.99 & 6576    &         15.13 &  	    56.0 &  5140 &   207 & 2.62 &     0.0\\
1820    &         12.76 &              86.0 &  4479 &	333 & 2.58 &  -0.6 & 6808    &         22.54 &  	    97.0 &  6652 &   -99 & \nodata &	5.91\\
2182	&	  14.66	&	       78.0 &  5920 &	187 & 3.35 &   3.0 & 6811    &         17.41 &  	    90.0 &  5952 &   137 & 3.07 &    3.65\\
2456	&	  14.82	&	       48.0 &  4809 &	301 & 2.85 &  -0.15& 6974    &         15.38 &  	     5.3 &  4625 &    74 & 1.16 &    0.9\\
2457	&	  -7.35	&	      177.0 &  4527 &	-99 & \nodata &0.0 & 6978    &         15.05 &  	     7.4 &  4668 &   179 & 1.84 &    0.17\\
2540	&	  15.14	&	        6.2 &  3576 &	  0 & \nodata &-1.8& 7372    &         15.80 &  	     1.8 &  4461 &    39 & 0.60 &    0.53\\ 
2550	&	  -1.75	&	        2.6 &  5053 &	 0 & \nodata  &1.6 & 7422    &         14.83 &  	    33.0 &  5651 &   233 & 3.40 &    2.47\\
2578	&	  27.78	&	        2.9 &  4694 &	  0 & \nodata &0.7 & 7442    &        -19.82 &  	     2.3 &  4591 &     0 & \nodata &	0.8 \\
2717	&	  14.96	&	        3.4 &  5267 &	 24 & 1.41 &   3.4 & 7451    &         18.22 &  	     1.0 &  4519 &     0 & \nodata &	1.60\\
3359	&	  14.69	&	        8.6 &  5280 &	226 & 2.93 &  1.48 & 7663    &         25.70 &  	     5.4 &  5475 &    49 & 1.96 &    2.8 \\
3497	&	  -5.24	&	       14.4 &  5622 &	126 & 2.68 &  2.66 & 7670    &         41.08 &  	     4.1 &  5281 &    63 & 1.89 &    2.3 \\
3695	&	  15.16	&	       90.0 &  3287 &	-99 & \nodata&-4.6 & 7711    &         73.58 &  	     4.0 &  5097 &     0 & \nodata &	1.77\\
4280	&	  12.26	&	       16.0 &  5729 &	151 & 2.94 &  3.05 & 7857    &         17.95 &  	     7.7 &  6072 &   114 & 3.03 &    2.82\\
4324	&	  14.31	&	       41.0 &  6334 &	 97 & 3.14 &  4.27 & 7956    &         15.29 &  	    13.8 &  4759 &   289 & 2.69 &   -0.38\\
4362	&	  15.11	&	        9.5 &  5616 &	191 & 3.08 &  2.39 & 7973    &         18.81 &  	     2.0 &  4730 &    10 & 0.35 &    1.71\\
4413	&	  14.48	&	       11.4 &  5615 &	163 & 2.90 &  2.34 & 7974    &         19.80 &  	     7.5 &  5985 &    62 & 2.57 &    2.98\\
4454	&	  16.54	&	       93.5 &  6761 &	 83 & 3.35 &  6.61 & &&&&&&\\
\hline
\end{tabular}
\end{table*}

\newpage
\setcounter{table}{8}
\begin{table*}[htb]
\caption[]{\label{tab:patten}Cross-identifications with\\
 Patten \& Simon (1996)}
\setlength{\tabcolsep}{1mm}
\begin{tabular}{lcr|lcr}\hline\hline
VXR & No. & $P_{\mu}$ & VXR & No. & $P_{\mu}$ \\ \hline

   2b  & 4336 & 73 &    45a & 4658 & 75\\
   3a  & 4362 & 67 &    46  & 3696 & 94\\
   4   & 3445 & 81 &    47  & 3695 & 80\\
   5   & 4413 & 75 &    48  & 3709 & 86\\
   6a  & 2540 & 78 &    49b & 4695 & 46\\
   7   & 3464 & 72 &    50a & 6576 &  0\\
   8   & 5459 & 25 &    52  & 3722 & 84\\
   11  & 3497 & 63 &    54a & 5768 &  0\\
   12  & 4467 & 82 &    56  & 3746 & 53\\
   13  & 2606 & 86 &    57a & 6625 &  0\\
   14  & 1083 & 85 &    60a & 4757 & 28\\
   15  & 1715 &  0 &    62a & 5829 & 42\\
   16a & 8415 & 78 &    65  & 4778 & 39\\
   18  & 1759 & 80 &    66  & 2888 & 82\\
   21  & 4522 & 89 &    67a & 5859 & 78\\
   22a & 1142 & 73 &    69a & 2012 & 78\\
   24b & 4543 &  0 &    70  & 4809 & 33\\
   27  & 2672 &  0 &    72  & 5884 & 78\\
   30  & 4549 & 83 &    73  & 3900 & 59\\
   31  & 6478 &  0 &    75a & 4888 &  0\\
   35a & 1820 &  9 &    76a & 4902 & 78\\
   38a & 3649 &  0 &    77a & 6811 & 81\\
   41  & 4636 & 75 &    78  & 2188 &  0\\
   43  & 5696 &  0 &    79a & 5050 &  0\\
   44  & 3683 & 62 &        &      &   \\
\hline
\end{tabular}
\end{table*}

\newpage
\setcounter{table}{9}
\begin{table*}[htb]
\caption[]{\label{tab:param}Orbital elements of 4413 = VXR 5}
\begin{tabular}{lll}\hline\hline
Element (units) & Value & $\sigma$  \\ \hline

P (d) & 90.617 & 0.007\\
T (JD-2400000) & 45025.79 & 0.51\\
e              & 0.287 & 0.007\\
$\gamma$-velocity (km s$^{-1}$) & 14.35 & 0.09\\
$\omega$ (\hbox{$^\circ$}) & 22.8 & 1.3\\
K$_1$ (km s$^{-1}$) & 29.51 & 0.25\\
K$_2$ (km s$^{-1}$) & 30.54 & 0.26\\
$a_1 \sin i$ (Gm) & 35.23 & 0.38\\
$a_2 \sin i$ (Gm) & 36.45 & 0.39\\
$\sigma$(O-C) (km s$^{-1}$) & 0.60 & \nodata\\
n$_{\rm obs}$ & 22 & \nodata\\
\hline
\end{tabular}
\end{table*}

\end{document}